\documentclass[epj,nopacs]{svjour}
\usepackage{epsfig}
\usepackage{latexsym}
\usepackage{graphics}
 
\newcommand{\be}{\begin{equation}}
\newcommand{\ee}{\end{equation}}
\newcommand{\bear}{\be\begin{array}}
\newcommand{\bea}{\begin{eqnarray}}
\newcommand{\eea}{\end{eqnarray}}
\newcommand{\dst}{\displaystyle}
\newcommand{\fr}[2]{\frac{{\dst #1}}{{\dst #2}}}
\newcommand{\ggam}{\mbox{$\gamma\gamma\,$}}
\newcommand{\bm}{\boldmath}
\newcommand{\epe}{\mbox{$e^+e^-\,$}}

\newcommand{\fn }[1]{\footnote{ #1}}

\def\lsi{\raise0.3ex\hbox{$<$\kern-0.75em\raise-1.1ex\hbox{$\sim$}}}
\def\gsi{\raise0.3ex\hbox{$>$\kern-0.75em\raise-1.1ex\hbox{$\sim$}}}
\newcommand{\lsim}{\mathop{\lsi}}
\newcommand{\gsim}{\mathop{\gsi}}
\newcommand\vep{\varepsilon}

\newcounter{saveeqn}
\newcommand{\appeqn}{\setcounter{saveeqn}{\value{equation}}%
\setcounter{equation}{0}%
\renewcommand{\theequation}{%
\mbox{\Alph{section}.\arabic{equation}}}}%
\newcommand{\reseteqn}{\setcounter{equation}{\value{saveeqn}}%
\renewcommand{\theequation}{\arabic{equation}}}

\journalname{EPJ C}
\begin{document}
\setcounter{page}{1}
\title{
  \vspace{-23pt}
  {\rm \rightline{\small LU 2000/024}}
  Charge asymmetry of pions in the process $e^-e^+\to e^-e^+\pi^+\pi^-$
}
\titlerunning{Charge asymmetry of pions
              in the process $e^-e^+\to e^-e^+\pi^+\pi^-$}
\authorrunning{I.F.~Ginzburg et al.}
\author{I.F.~Ginzburg \inst{1,2,}\thanks{ginzburg@math.nsc.ru} %
\and A.~Schiller \inst{3,}\thanks{Arwed.Schiller@itp.uni-leipzig.de} %
\and V.G.~Serbo \inst{2,}\thanks{serbo@math.nsc.ru}}
\institute{Sobolev Institute of Mathematics, Novosibirsk, 630090
Russia \and Novosibirsk State University, Novosibirsk, 630090
Russia \and Institut f\"ur Theoretische Physik and NTZ,
Universit\"at Leipzig, D-04109 Leipzig, F.R.~Germany}
\date{Received: December 7, 2000}
\abstract{
  The study of the charge asymmetry of produced particles allows to investigate
  the interference of different production mechanisms and to  determine new
  features of the corresponding amplitudes. In the process
  $e^- e^+ \to e^- e^+ \pi^+ \pi^-$ the two--pion system is produced via two
  mechanisms: two--photon (C--even state) and bremsstrahlung (C--odd state)
  production. We study the charge asymmetry of pions in a  differential in the
  pion momenta cross section originating from an interference between these
  two mechanisms. At low effective mass of dipions this asymmetry is directly
  related to the $s$-- and $p$--phases of elastic $\pi\pi$ scattering. At
  higher energies it can give new information about the $f_0$ meson family,
  $f_2(1270)$ meson, etc. The asymmetry is expressed via the pion form factor
  $F_\pi$ and helicity amplitudes $M_{ab}$ for the subprocess
  $\gamma^*\gamma\to \pi^+\pi^-$ as $\sum G_{ab}{\rm Re}(F_\pi^*M_{ab})$
  where we have calculated analytically the coefficients $G_{ab}$ for the
  region giving the main contribution to the effect. Several distributions of
  pions are presented performing a numerical analysis in a model with
  point--like pions. In the region near the dipion threshold the asymmetry is
  of the order of $1 \%$. We show that with suitable cuts the signal to
  background ratio can be increased up to about 10\%.
}

\maketitle

\section{Introduction}

\begin{figure}[!htb]
  \centering
  \unitlength=1.55mm
    \begin{picture}(42.00,20.00)
    \put(19.00,15.00){\makebox(0,0)[cc]{$q_1$}}
    \put(19.00,7.00){\makebox(0,0)[cc]{$q_2$}}
    \put(21.00,0.01){\makebox(0,0)[c]{${\cal M}_1$}}
    \put(35.00,18.80){\makebox(0,0)[r]{$p'_1$}}
    \put(5.00,18.80){\makebox(0,0)[l]{$p_1$}}
    \put(35.00,3.20){\makebox(0,0)[r]{$-p'_2$}}
    \put(5.00,3.20){\makebox(0,0)[l]{$-p_2$}}
    \put(21.00,5.00){\line(0,1){1.00}}
    \put(21.00,9.50){\line(0,1){1.00}}
    \put(21.00,17.00){\line(0,-1){1.00}}
    \put(21.00,12.50){\line(0,-1){1.00}}
    \put(21.00,11.00){\circle*{1.50}}
    \put(20.90,11.25){\line(2,1){4.00}}
    \put(20.70,11.45){\line(2,1){4.00}}
    \put(22.50,10.15){\line(2,-1){4.00}}
    \put(22.30,10.00){\line(2,-1){4.00}}
    \put(21.60,10.50){\vector(-2,1){0.00}}
    \put(25.70,13.80){\vector(2,1){0.00}}
    \put(27.70,13.80){\makebox(0,0)[cc]{$p_-$}}
    \put(28.70,8.00){\makebox(0,0)[cc]{$-p_+$}}
    \put(21.00,15.00){\vector(0,-1){1.50}}
    \put(21.00,7.00){\vector(0,1){1.50}}
    \put(05.00,17.00){\vector(1,0){5.00}}
    \put(10.00,17.00){\vector(1,0){25.00}}
    \put(35.00,5.00){\vector(-1,0){5.00}}
    \put(30.00,5.00){\vector(-1,0){25.00}}
  \end{picture}
  \caption{
           Amplitude ${\cal M}_1$ for the two--photon production
           of pions. The $e^-$ and $e^+$ with initial 4--momenta
           (energies) $p_1$ ($E_1$) and $p_2$ ($E_2$) and final momenta
           (energies) $p^\prime_1$ ($E_1^\prime$) and $p^\prime_2$
           ($E_2^\prime$) emit virtual photons with  $q_i =p_i -
           p_i^\prime$ ($\omega_i =E_i-E_i^\prime$).  These photons produce
           the $C$--even $\pi^+ \pi^-$ system with total 4--momentum $k=p_+
           +p_-=q_1+q_2$ and effective mass {$W=\protect\sqrt{k^2}$},
           furthermore $s=(p_1+p_2)^2 = 4 E_1 E_2$.
          }
  \label{fig:1}
\end{figure}
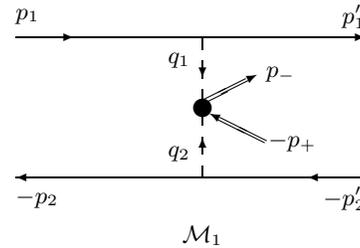

The study of charge asymmetry in  particle production can be used
as an essential source of information about production amplitudes
which is difficult to obtain otherwise. In this paper we discuss
the charge asymmetry of pions produced in the reaction $e^-e^+\to
e^-e^+\pi^+\pi^-$.  The pion pair (dipion) in this process is
produced mainly via the two--photon mechanism (Fig.~\ref{fig:1})
or via  bremsstrahlung (Fig.~\ref{fig:2}).
\begin{figure}[!htb]
  \centering
  \unitlength=1.55mm
  \begin{picture}(55.00,30.00)
    \put(25.00,11.00){\vector(-1,0){4.00}}
    \put(21.00,11.00){\vector(-1,0){20.00}}
    \put(21.30,25.70){\vector(2,1){0.00}}
    \put(16.50,23.40){\line(2,1){4.00}}
    \put(16.50,23.20){\line(2,1){4.00}}
    \put(17.20,22.50){\vector(-2,1){0.00}}
    \put(17.90,22.00){\line(2,-1){4.00}}
    \put(17.90,22.20){\line(2,-1){4.00}}
    \put(10.70,19.00){\vector(1,0){14.00}}
    \put(5.00,19.00){\line(1,0){2.20}}
    \put(10.50,23.00){\line(1,0){1.00}}
    \put(12.50,23.00){\line(1,0){1.00}}
    \put(14.50,23.00){\vector(1,0){1.50}}
    \put(9.00,21.00){\circle{4.60}}
    \put(1.00,19.00){\vector(1,0){4.00}}
    \put(16.6,23){\circle*{1.5}}
    \put(9.00,13.00){\vector(0,1){2.00}}
    \put(9.00,11.00){\line(0,1){1.00}}
    \put(9.00,15.70){\line(0,1){1.0}}
    \put(9.00,17.80){\line(0,1){0.8}}
    \put(2.00,12.30){\makebox(0,0)[cc]{$-p_2$}}
    \put(22.00,12.30){\makebox(0,0)[cc]{$-p'_2$}}
    \put(10.40,14.00){\makebox(0,0)[cc]{$q_2$}}
    \put(3.00,17.20){\makebox(0,0)[cc]{$p_1$}}
    \put(22.00,17.20){\makebox(0,0)[cc]{$p'_1$}}
    \put(13.50,21.50){\makebox(0,0)[cc]{$k$}}
    \put(9.00,02.00){\makebox(0,0)[cc]{${\cal M}_2$}}
    \put(39.00,02.00){\makebox(0,0)[cc]{${\cal M}_3$}}
    \put(55.00,11.00){\vector(-1,0){4.00}}
    \put(51.00,11.00){\line(-1,0){10.3}}
    \put(37.30,11.00){\vector(-1,0){7.00}}
    \put(51.10,9.70){\vector(2,1){0.00}}
    \put(46.40,7.40){\line(2,1){4.00}}
    \put(46.40,7.20){\line(2,1){4.00}}
    \put(47.10,6.50){\vector(-2,1){0.00}}
    \put(47.80,6.00){\line(2,-1){4.00}}
    \put(47.80,6.20){\line(2,-1){4.00}}
    \put(40.70,19.00){\vector(1,0){14.00}}
    \put(35.00,19.00){\line(1,0){7.00}}
    \put(40.50,7.00){\line(1,0){1.00}}
    \put(42.50,7.00){\line(1,0){1.00}}
    \put(44.50,7.00){\vector(1,0){1.50}}
    \put(39.00,9.00){\circle{4.60}}
    \put(31.00,19.00){\vector(1,0){4.00}}
    \put(46.5,7){\circle*{1.5}}
    \put(39.00,15.00){\vector(0,-1){2.00}}
    \put(39.00,11.40){\line(0,1){1.00}}
    \put(39.00,15.70){\line(0,1){1.0}}
    \put(39.00,17.80){\line(0,1){1.2}}
    \put(40.40,14.00){\makebox(0,0)[cc]{$q_1$}}
    \put(43.50,8.50){\makebox(0,0)[cc]{$k$}}
  \end{picture}
  \caption{
           Amplitudes ${\cal M}_2$ and ${\cal M}_3$ for the
           bremsstrahlung production of pions. The pion pair in $C$--odd
           state is produced by one virtual photon with 4--momentum $k =p_+
           + p_-$ emitted by the electron (${\cal M}_2$) or by the positron
           (${\cal M}_3$). The open circles represent the virtual Compton
           scattering shown in Fig.~\protect\ref{fig:3}.
          }
  \label{fig:2}
\end{figure}
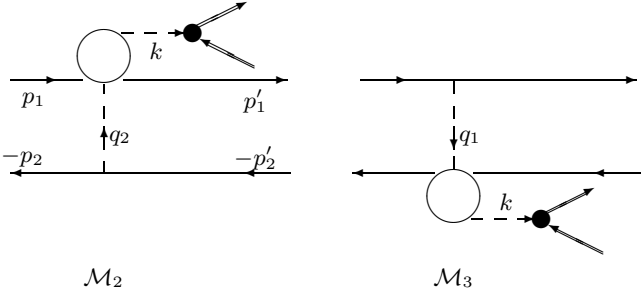
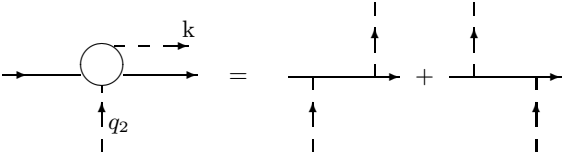
\begin{figure}[!htb]
  \centering
  \unitlength=1.65mm
  \begin{picture}(48.00,15.00)
    \put(2.00,9.20){\vector(1,0){2.00}}
    \put(10.00,10.00){\circle{3.40}}
    \put(3.00,9.20){\line(1,0){5.30}}
    \put(11.70,9.20){\vector(1,0){6.00}}
    \put(10.00,5.00){\vector(0,1){2.00}}
    \put(10.00,3.00){\line(0,1){1.00}}
    \put(10.00,7.80){\line(0,1){0.50}}
    \put(11.00,11.50){\line(1,0){0.80}}
    \put(13.00,11.50){\line(1,0){0.80}}
    \put(15.00,11.50){\vector(1,0){2.00}}
    \put(17.00,12.90){\makebox(0,0)[cc]{$$k$$}}
    \put(25.00,9.00){\vector(1,0){9.00}}
    \put(38.00,9.00){\vector(1,0){9.00}}
    \put(27.00,8.00){\line(0,1){1.00}}
    \put(27.00,5.00){\vector(0,1){2.00}}
    \put(27.00,3.00){\line(0,1){1.00}}
    \put(32.00,9.00){\line(0,1){1.00}}
    \put(32.00,11.00){\vector(0,1){2.00}}
    \put(32.00,14.00){\line(0,1){1.00}}
    \put(40.00,9.00){\line(0,1){1.00}}
    \put(40.00,11.00){\vector(0,1){2.00}}
    \put(40.00,14.00){\line(0,1){1.00}}
    \put(45.00,8.00){\line(0,1){1.00}}
    \put(45.00,5.00){\vector(0,1){2.00}}
    \put(45.00,3.00){\line(0,1){1.00}}
    \put(21.00,9.00){\makebox(0,0)[cc]{=}}
    \put(36.00,9.00){\makebox(0,0)[cc]{+}}
    \put(11.40,5.00){\makebox(0,0)[cc]{$q_2$}}
  \end{picture}
  \caption{Virtual Compton scattering.}
  \label{fig:3}
\end{figure}

The first of them gives charge parity even ($C$--even)
dipions, while the second
mechanism leads to $C$--odd pion pairs. In the differential
cross section the interference of these mechanisms results in
terms which are antisymmetric under $\pi^+\leftrightarrow \pi^-$
exchange. Certainly, these terms disappear in the total cross
section or after a suitable averaging. Nevertheless, they are
observable and could give new information about the production
amplitudes which cannot be obtained unambiguously by other
approaches. In particular, at low effective masses of dipions
$W\sim 2m_\pi$ this interference is directly related to the
difference of $s$-- and $p$--phase shifts of the elastic
$\pi\pi$ scattering. These phase shifts are of primary
importance for low energy hadron physics \cite{DAFNE,Phi} (in
particular, for the chiral dynamics which pretends to be the low
energy QCD). At higher energies the interference can give new
information about the $f_0(400-1200)$ meson (former $\sigma$),
the $f_0(980)$ meson, etc.  Their nature is now subject of
wide discussions.

The opportunity using $C$--odd effects for such problems was
firstly studied almost three decades ago in Ref.~\cite{CS}. In
that paper the case of the small total transverse momenta of the
produced pion pair, ${\bf k}_\perp^2 \ll m_\pi^2$, was considered.
However, this region gives only a small fraction of the entire
charge asymmetry discussed. In this paper we obtain formulae which
allow to study the charge asymmetry in the main kinematical
region, $k_\perp \lsim W$, and discuss the main features of
the effect and the background.

Recently the process $e^+e^-\to e^+e^-\pi^+\pi^-$ was considered
in Ref.~\cite{Diehl} for the case when the virtuality of one
photon is large, $-q_1^2\gg W^2$. The authors concentrate their efforts
on a QCD analysis of the exclusive dipion production in
$\gamma^*\gamma$ collision in that region (leading to a
factorization of perturbative QCD subprocesses and a generalized
two--meson distribution amplitude). The developed description of
the process at $e^+e^-$ colliders also includes the interference
of two--photon and bremsstrahlung production mechanisms.
Naturally, the estimates for possible number of events of that
work are considerably lower than those in the main region
considered in the present paper. Note that the discussed charge
asymmetry was observed at CLEO in the $\epe\to\epe\pi\pi$
reaction detecting additionally an electron scattered at large
angle~\cite{Sav}\fn{A similar problem was discussed in
Ref.~\cite{Kur} for $ep$--scattering related to HERA
experiments. Unfortunately, those results have no direct
relation to experiments since the main $C$--odd contribution in
$ep$--scattering is given by the production of $\rho$ mesons via
strong interactions (diffractive production for pions flying
along electrons or proton excitation for pions flying along
protons). The bremsstrahlung production considered in \cite{Kur}
is suppressed  roughly by a factor $\alpha =1/137$.  Besides,
the authors of Ref.~\cite{Kur} claim that their formulae
(obtained for $ep \to ep \pi^+ \pi^-$) are valid for the pion
charge asymmetry in  the process $e^-e^+\to e^-e^+ \pi^+\pi^-$. In
that respect their results are definitely incorrect since
contributions of zero helicity  for the virtual photon are not
included. As we show here, these contributions are of the
same order of magnitude  as those with helicity $\pm 1$ (see
Sect.~\ref{sec:pi} for details).}.

The effect of interest can be studied on $e^+e^-$ colliders with
c.m. energies above $1$ GeV. We show that the charge asymmetry is
of the order of $1 \%$ and that the signal to background ratio can
be considerably improved with suitable cuts. Let us emphasize
that in the method of data preparation suggested in the present
paper, the considered asymmetries can be obtained from the data
independent on the uncertainty in calculating the background.

We consider an experimental set--up when only pion momenta ${\bf
p_+}$ and ${\bf p_-}$ are measured (the so called {\em no tag}
experiments). This set--up corresponds to the cross section $d\sigma/
(d^3p_+ d^3p_-)$ for the $e^- e^+ \to e^- e^+ \pi^+ \pi^-$
process. Besides, our results are also valid for the {\em  single
tag} experiments in which additionally the scattered electron is
recorded, and where an averaging is performed over the small
unbalance of transverse momentum of the scattered electron and
the dipion.

The {\bf outline} of our paper is as follows: First we present a
qualitative description of different contributions to the reaction
$e^-e^+ \to e^-e^+\pi^+\pi^-$.  In Sect.~\ref{sec:bas} the necessary
variables are defined and the basic formulae are presented.  The
amplitude of the subprocess $\gamma^*\gamma^* \to \pi^+\pi^-$ is
represented via helicity amplitudes in a model independent way.
The charge asymmetry of pions is calculated in Sect.~\ref{sec:pi}.
The obtained result [see Eqs.~(\ref{7}),(\ref{35}) and (\ref{23})]
is given in a simple
analytical form. To discuss the background problems more
accurately, we present in Sect.~\ref{sec:bg} approximate formulae
for the two--photon and bremsstrahlung production in the
kinematical region which is essential for the charge asymmetry. In
Sect.~\ref{sec:ggpipi} we present an approximate description of
the $\gamma^* \gamma \to \pi^+\pi^-$
subprocess entering the two--photon amplitude.
To get an idea about the potentiality
of future experiments we perform
a numerical analysis in Sect.~\ref{sec:analysis}
restricting ourselves to the QED model (point--like pions)
for the amplitudes which gives a
reasonable description at $W < 1$ GeV.
We present several important distributions and estimate the
background. Additionally we study the charge asymmetry of muons in the
process $\epe \to \epe\mu^+\mu^-$ (Sect.~\ref{sec:muon}). In the
final Section we summarize our results. Details of the
calculations are presented in the Appendices.


\section{Qualitative description of different contributions}
\label{sec:qual}


The main contribution to the cross section of the process
can be written via the amplitudes ${\cal M}_j$ shown in
Figs.~\ref{fig:1},\ref{fig:2}. It is a sum of $C$--even, $C$--
odd and interference contributions:
\be
  d\sigma = d\sigma_{C=+1}+ d\sigma_{C=-1} +d\sigma_{\rm interf} \,,
  \label{3}
\ee
where
\bea
  &&d\sigma_{C=+1} \propto |\,{\cal  M}_1 \,|^2\,,
  \nonumber \\
  &&d\sigma_{C=-1}= d\sigma_2 + d\sigma_3
  \propto |\,{\cal M}_2|^2 \,+\, |{\cal M}_3 \,|^2\,,
  \nonumber \\
  &&d\sigma_{\rm interf}=d\sigma_{12}+d\sigma_{13}\,,
  \label{5}
  \\
  &&d\sigma_{12} \propto 2\,{\rm Re}({\cal M}_2^* {\cal M}_1) \,, \quad
  d\sigma_{13} \propto 2\,{\rm Re}({\cal M}_3^* {\cal M}_1)\,.
  \nonumber
\eea
Let us discuss these contributions qualitatively.  In the
head--on collisions of the leptons the $z$--axis is chosen along
the initial electron momentum.

{\bf The two--photon mechanism} (Fig.~\ref{fig:1}) produces
$C$--even dipions. It provides the main contribution to the
total cross section. The corresponding part of the cross section
$d\sigma_{C=+1}$ can be expressed via the amplitudes $M_{ab}$
describing the collisions of virtual photons with helicities $a$
and $b$ ($a,b=\pm 1,\;0$) \cite{BG}. Its dominant part is given
by almost real photons which have virtualities $q_1^2$ and
$q_2^2$ close to zero (small transfer momenta squared of
electrons and positrons). The produced pairs are distributed
almost uniformly over their total rapidity and peaked at small
values of their total transverse momentum $k_\perp$ (for details
see review~\cite{BGMS}). In this kinematical region only the
transverse  helicities ($a,b=\pm 1$)  for almost on-shell
photons give the dominant contribution and the cross section can
be written via $|M_{++}|^2$, $|M_{+-}|^2$ and ${\rm
Re}(M^*_{+-}M_{++})$.

{\bf The bremsstrahlung mechanism} (Fig.~\ref{fig:2}) produces
pion pairs in $C$--odd state. Its contribution to the cross
section $d\sigma_{C=-1}/(d^3p_+\, d^3p_-)$ was calculated in
Ref.~\cite{KS}. It is proportional to $|F_\pi(k^2)|^2$ where
$F_\pi$ is the pion form factor. The main contribution to
$d\sigma_2$ is given by the region where the exchange photon
with momentum $q_2$ is almost real. In that domain the sum of
energy and longitudinal momentum of the produced pair is close
to that of the initial electron, while the transverse momentum
of the pair $k_\perp$ is very small.

However, in the kinematical region being essential for both
two--photon production and interference, the
$k_\perp$--distribution of the pions is relatively wide. The
interference between bremsstrahlung by an electron (amplitude
${\cal M}_2$) and by a positron (${\cal M}_3$) is negligible
small. Note that both contributions $d\sigma_{C=+1}$ and
$d\sigma_{C=-1}$ are charge symmetric, they do not change under
pion exchange $\pi^+ \leftrightarrow \pi^-$.

{\bf The interference of $C$--even and $C$--odd contributions}
$d\sigma_{\rm interf}=d\sigma_{12}+ d\sigma_{13}$ [see
Eq.~(\ref{5})] is antisymmetric under  pion exchange $\pi^+
\leftrightarrow \pi^-$ due to opposite charge parities of dipion
states produced by two--photon and bremsstrahlung mechanisms.
Therefore, this interference determine {\em the charge asymmetry
of pions}, i.e.  \be d\sigma_{\rm interf}=
  \fr{1}{2}\,\left[\,d\sigma(p_+,p_-,\dots)-d\sigma(p_-,p_+,\dots)
  \right]\,.
  \label{6}
\ee
To discuss that asymmetry, it is useful to
introduce the operator of charge conjugation of pions $\hat C_\pi$
by its action on an arbitrary function $F(p_+,p_-)$:
\be
  \hat{C_\pi}F(p_+,p_-)=F(p_-,p_+)\,.
  \label{defc}
\ee
In particular, we have
$$
  \hat{C_\pi} d\sigma_{C=\pm 1} = d\sigma_{C=\pm 1}\,,\quad
  \hat{C_\pi} d\sigma_{\rm interf}=-d\sigma_{\rm interf}\,.
$$

Many features of this interference can be naturally explained
taking into account the described features of the two--photon
and bremsstrahlung production. For example, the main
contribution to $d\sigma_{12}$ is given by an almost real photon
$q_2$ (small transfer momentum squared of the positron). The
produced pions fly mainly along the electron, $k_z=p_{+z}
+p_{-z}>0$, and the dipion transverse momentum distribution is
not peaked at small $k_\perp$. Therefore the transverse momentum
of the electron is not small, ${\bf q}_{1\perp}\approx {\bf
k}_\perp$.
Similarly, the main contribution to $d\sigma_{13}$ is given by
the almost real photon $q_1$ (small transfer momentum squared of
the electron), whereas the transverse momentum of the positron
is not small, ${\bf q}_{2\perp} \approx {\bf k}_\perp$. The
produced pions fly mainly along the positron, $k_z<0$.

Certainly, the sign of the observed effect is different
for $d\sigma_{12}$ (related to brems\-strah\-lung production of
pions by an electron with negative charge) and
for $d\sigma_{13}$ (related to brems\-strah\-lung production of
pions by an positron with positive charge).
Therefore, some details of the asymmetry are different for $e^+e^-$
and $e^-e^-$ collisions.

{\bf The $s$--channel contributions}. In the previous discussion
we did not consider the additional set of diagrams which can be
obtained from those in Figs.~\ref{fig:1},\ref{fig:2}
interchanging the outgoing electron by the incoming positron
($p_1'\leftrightarrow - p_2$), etc.  The contributions of those
diagrams contain an additional factor $1/s$ due to the photon
propagator.  Besides, the final electron and positron have a
wide angular distribution in the main regions   and, therefore,
do not give a logarithmic enhancement (contrary to the
considered diagrams).  As a result, the contributions of the
$s$--channel (annihilation) diagrams and their interference with
those of Figs.~\ref{fig:1},~\ref{fig:2} are suppressed by a
factor $W^2/(sL)$ where $L\sim 10\div 15$ is a typical
 logarithm.

\section{Basic notations and general formulae}
\label{sec:bas}

The main notations are given in Figs.~\ref{fig:1} and
\ref{fig:2}. As already mentioned, we consider the head--on
collisions of electrons and positrons with the $z$--axis along
the initial electron momentum.  In this frame $p_i=
(E_i,\,0,\,0,\,\pm \sqrt{E_i^2- m^2_e})$, $i=1,2$. The virtual
photon momenta for the two--photon production of
Fig.~\ref{fig:1} are $q_i =p_i- p_i^\prime = (\omega_i,\, {\bf
q}_{i\perp},\, q_{iz})$ with
\be
  q^2_i = 2 q_ip_i =\,-\, \fr{{\bf q}_{i\perp}^2 +
  m_e^2\,(\omega_i/E_i)^2 }{ 1-(\omega_i/E_i)} \,<\, 0\,.
  \label{9}
\ee
The 4--momenta of the produced pions are given by $p_{\pm} =(\vep_{\pm},\,
{\bf p}_{\pm\perp},\, p_{\pm z})$ with $p^2_{\pm} = \mu^2$.
Below we use the quantities
\bea
  &&x_{\pm} =\fr{\varepsilon_{\pm} + p_{\pm z}}{ 2E_1 }=\fr{p_{\pm}p_2
  }{ p_1p_2} \,,\quad
  y_{\pm} = \fr{\varepsilon_{\pm} - p_{\pm z}}{ 2E_2}
  =\fr{p_{\pm}p_1}{ p_2p_1} \,,
  \nonumber
  \\
  &&x=x_+ + x_-\,, \quad y=y_+ + y_-\,.
  \label{12}
\eea
For ultra--relativistic pions flying along the initial electron
or positron momentum, the quantity $x_\pm$ ($y_\pm$) is the
fraction of energy transferred from the electron (positron) to
$\pi^\pm$. The variables $x_\pm$ ($y_\pm$) appear in the cross
section $d\sigma_{12}$ ($d\sigma_{13}$).

In what follows, we consider the symmetric and antisymmetric combinations
of the pion momenta:
\be
  k= p_+ +p_-\,, \qquad \Delta = p_+ -p_-\,.
  \label{12a}
\ee
The pion charge conjugation operator leads in particular to
$$
  \hat{C_\pi}\Delta^\mu=-\Delta^\mu\,.
$$
Therefore, the asymmetry effects are proportional to the components of the
4--vector $\Delta$.

To describe that asymmetry we use the variables
\bea
\xi &=&\fr{x_+ -x_-}{x} =
  \fr{p_2\,\Delta}{p_2k}\,,
  \nonumber \\
  \eta&=&\fr{y_+-y_-}{y}=\fr{p_1\,\Delta}{p_1k}\,,
  \label{13}
  \\
  K_-&=&\fr{(p_2-p_1)\,\Delta}{(p_2+p_1)\,k}=\fr{x\xi-y\eta}{x+y}\,,
  \nonumber \\
  v&=& {\bf p}_{+\perp}^2 -
  {\bf p}_{-\perp}^2 = {\bf k}_{\perp} {\bf \Delta}_{\perp}\,.
  \nonumber
\eea
The ``transverse'' variable $v$ is a natural variable  both for
contributions $d\sigma_{12}$ and $d\sigma_{13}$. The
``longitudinal'' variable $\xi$ naturally arises in describing
the contribution $d\sigma_{12}$ (whereas $\eta$ -- in
describing $d\sigma_{13}$). The symmetric variable $K_-$ is
suitable to discuss the sum $d\sigma_{12} +d\sigma_{13}$.  Note
that $K_-$ is proportional to the difference of the longitudinal
momenta of $\pi^+$ and $\pi^-$ in the $e^- e^+$ center--of--mass
system
$$
K_-= \left\{\fr{p_{+z}-p_{-z}}{\vep_+ +\vep_-}\right\}_{e^-e^+
  \,{\mathrm{c.m.s.}}} \,.
$$
Besides, $K_-=\xi$ at $x\gg y$ and $K_-=-\eta$ at $x\ll y$.

{\bf\bm The amplitude ${\cal M}_1$ of the two--photon
production} is written via the amplitude $M^{\mu \nu}$ of the
subprocess $\gamma^* \gamma^* \to \pi^+ \pi^-$ as
\be
  {\cal M}_1 = \fr{4\pi \alpha }{ q_1^2 q_2^2}\,
  \left(\bar u_1^\prime \gamma_\mu u_1\right)\,
  \left(\bar v_2 \gamma_\nu v_2^\prime \right)\, M^{\mu \nu}
  \label{15}
\ee
where the bispinors $u_1 \,(u_1^\prime)$ and $v_2 \,
(v_2^\prime)$ correspond to the initial (final) electron and
positron, respectively.

Instead of the amplitude $M^{\mu \nu}$  it is more
convenient to use the helicity amplitudes $M_{ab}$ which can be
introduced via\fn{Details of kinematics for the subprocess
$\gamma^* \gamma^* \to \pi^+ \pi^-$ are given in Appendix~\ref{app:A}.}
\be
  M_{\mu \nu}=\sum\limits_{ab=\pm 1, \,0}\,(-1)^{a+b}
  e_{1\mu}^{(a)*}\, e_{2\nu}^{(b)*}\, M_{ab}\,.
  \label{16}
\ee
Here $e_{j\mu}^{(a)}$ is the polarization vector of the $j$--th
virtual photon with helicity $a=\pm 1,\,0$. A virtual photon is
called {\it transverse} if its helicity is equal to $\pm 1$ and
{\it scalar (longitudinal)} for zero helicity.
Since the amplitude $M^{\mu\nu}$ is C--even, the vectors of
transverse and scalar polarization are C--odd and C--even, respectively
(see Appendix~\ref{app:A} for details):
\bea
  \hat{C_\pi}M^{\mu\nu}&=&M^{\mu\nu}\,,
  \nonumber \\
  \hat{C_\pi}e^{(\pm 1)}_{i\mu} = -e^{(\pm 1)}_{i\mu}&,& \quad
  \hat{C_\pi}e^{(0)}_{i\mu}=e^{(0)}_{i\mu}\,,
  \label{Cact}
\eea
we get
\be
  \hat{C_\pi}M_{0\pm}=-M_{0\pm}\,,\quad
  \hat{C_\pi}M_{+\pm}=M_{+\pm}\,.
  \label{C2prop}
\ee
It is important that the amplitudes $M_{ab}$ with scalar photons
disappear near the photon mass shell
\bea
  M_{0\pm} \propto \sqrt{-q_1^2}\,, \quad M_{\pm 0} &\propto&
  \sqrt{-q_2^2}\,, \quad M_{0 0} \propto \sqrt{q_1^2 q_2^2}
  \nonumber \\
  {\mathrm {at}} \;\; q_{1,2}^2 &\to& 0\,.
 \label{24}
\eea

So the amplitude ${\cal M}_1$ of the two--photon production can
be represented in the form
\bea
  {\cal M}_1 &=& \fr{4\pi \alpha }{ q_1^2 q_2^2}\,
  \sum\limits_{ab=\pm 1, \,0}
   (-1)^{a+b} \times
   \nonumber \\
  &&\left(\bar u_1^\prime \, \hat e_{1}^{(a)*}\, u_1\right)
               \left(\bar v_2  \, \hat e_{2}^{(b)*} \, v_2^\prime \right)
               M_{ab}\,.
  \label{17}
\eea

In a similar way {\bf\bm the amplitude ${\cal M}_2$ of the
brems\-strah\-lung production} by an electron  can be written as
\bea
  {\cal M}_2 &=& \fr{(4\pi  \alpha)^2}{ q_2^2 k^2}\, {F_\pi (k^2)}\times
  \nonumber \\
  &&\sum\limits_{c=\pm 1, \,0}\, (-1)^c \, \left(\bar u_1^\prime \,
  \hat{C}^{(c)}\, \, u_1\right)\, \left(\bar v_2 \,\hat
  e_{2}^{(c)*}\, v_2^\prime \right)\,,
  \label{18}
  \\
  \hat{C}^{(c)}&=& \hat
  \Delta \, \fr{\hat p_1^\prime +\hat k +m_e}{ (p_1^\prime +k)^2
  -m_e^2}\, \hat e_2^{(c)}+ \hat  e_2^{(c)}  \,\fr {\hat p_1 -\hat
  k +m_e}{ (p_1-k)^2 -m_e^2}\, \hat \Delta
  \nonumber
\eea
where the quantity $\bar u_1^\prime \, \hat{C}^{(c)}\, u_1$
corresponds to the amplitude of the virtual Compton scattering
shown in Fig.~\ref{fig:3}, $F_\pi(k^2)$ is the pion form
factor.

As a result, {\bf\bm the interference of the amplitudes of
${\cal M}_1$ and ${\cal M}_2$} is given by the expression
\bea
  d\sigma_{12} &=& 2\,{\rm Re}({\cal M}_2^* {\cal
  M}_1) \, \fr{d\Gamma }{ 2s}
  \label{19}
  \\
    &=& -2\, \fr{(4\pi \alpha)^3 }{ q_1^2
  q_2^2 k^2}\, \sum\limits_{abc=\pm 1, \,0}\! \! {\rm Re} \,\left(
  F_\pi^* \, M_{ab}\, \varrho_2^{bc}\, C_1^{ac}\, \right) \,
  \fr{d\Gamma}{ 2s}\,,
  \nonumber
\eea
where  the phase volume of final particles is
\bea
  d\Gamma &=&
  (2\pi)^4\,\delta(p_1+p_2-p_1^\prime -p_2^\prime -p_+ -p_-)\times
  \nonumber
  \\
  &&\fr{d^3 p_1^\prime\, d^3 p_2^\prime}{ 2E_1^\prime 2E_2^\prime
  \,(2\pi)^6}\, \fr{d^3 p_+\, d^3 p_+}{ 2\varepsilon_+
  2\varepsilon_- \,(2\pi)^6}\,,
  \label{20}
\eea
the (non-normalized) density matrix of the second virtual photon is
\be
  \rho_2^{bc}=\fr{(-1)^{b+c}}{2(-q_2^2)}{\mathrm{Tr}}\left\{
  (\hat{p}_2-m_e)\,\hat{e}_2^{(b)*}(\hat{p}_2^\prime-m_e)
  \,\hat{e}_2^{(c)} \right\}
  \label{21}
\ee
and
\be
  C_1^{ac}=\fr{(-1)^{a}}{2}{\mathrm{Tr}}\left\{
  (\hat{p}_1^\prime+m_e)\,\hat{e}_1^{(a)*}(\hat{p}_1+m_e)\,
  \hat{C}^{(c)*}
  \right\}\,.
  \label{22}
\ee

\section{The charge asymmetry}
\label{sec:pi}

Let us remind the reader
 that the charge asymmetry is determined by the
two
contributions  $d\sigma_{12}$ and $d\sigma_{13}$  arising from
the
interference of the two--photon diagram of Fig.~\ref{fig:1} with bremsstrahlung
diagrams of Fig.~\ref{fig:2}, [see Eq.~(\ref{5})]. Calculating that asymmetry
we limit ourselves to logarithmic accuracy (which is about 5 \% in our
case).

First, we discuss {\bf\bm the contribution
$d\sigma_{12}$}. According to the qualitative description in
Sect.~\ref{sec:qual} the main contribution to  $d\sigma_{12}$ is
given by very small values of $(-q_2^2)$. Therefore, the second
virtual photon can be considered as almost real.  Taking into
account Eq.~(\ref{9}) we can use in all expressions
\be
  q_2^2 =0\,,\;\;{\bf q}_{2\perp}=0\,,\quad
  {\bf q}_{1\perp} ={\bf k}_{\perp}\,,\quad
  q_1^2 = -\,\fr{{\bf k}_{\perp}^2}{ 1-x} \,,
  \label{26}
\ee
except for the propagator of that second photon
in the matrix element ${\cal M}_1$ [Eqs.~(\ref{15}),(\ref{17})].
This has the consequence that in that limit
the amplitudes $M_{ab}$ with $b=0$
can be safely neglected [see Eq.~(\ref{24})].

To obtain  $d\sigma_{12}/(d^3p_+ d^3p_-)$, we transform the
phase volume (\ref{20}) to the form
\be
  d\Gamma =\fr{d^2 q_{2\perp}}{ 32 (2\pi)^8\,(E_1-\omega_1)
  (E_2-\omega_2)}\, \fr{d^3 p_+\, d^3 p_+}{\varepsilon_+ \varepsilon_-}
  \label{25}
\ee
and integrate  either over ${\bf q}_{2\perp}$ or $q_2^2$ and
$\varphi_2$. The latter is the azimuthal angle of vector ${\bf
q}_{2\perp}$. After integrating over $\varphi_2$ the
non--diagonal elements of the $\rho_2^{bc}$ matrix disappear and
the final result contains $\rho_2^{++}= \rho_2^{--}$ only (i.e.
$b=c=\pm 1$)
$$
  \int\, \rho_2^{bc}\, \fr{d|q_2^2|}{ |q_2^2|}\,d\varphi_2
  = 2\pi\,\delta_{bc}\, \rho_2^{++}\, L_2\,,
$$
$$
  L_2 = \int\, \fr{ d|q_2^2|}{ |q_2^2|}=
  \ln\fr{|q_2^2|_{\max}}{|q_2^2|_{\min}}\,.
$$
In the integration over $|q_2^2|$ the lower limit is of
kinematical origin, $ |q_2^2|_{\min} = m_e^2 y_2^2/(1-y_2)$,
where
\be
  y_2 = \fr{2 q_2 p_1}{s}=
  \fr{\omega_2}{E_2} = \fr{W^2 (1-x) +{\bf k}_\perp^2}{s x(1-x)}\,.
  \label{28}
\ee
With the considered logarithmic accuracy the upper limit is
determined by a scale at which the integrand (besides the photon
propagator) starts to decrease significantly. For the pion pair
production this leads to
\be
  |q_2^2|_{\max}
  \sim \min \left\{\fr{{\bf k}_\perp^2}{1-y_2}, \,
  m_\rho^2, \, W^2 \right\}
  \label{qmax}
\ee
where $m_\rho$ is the $\rho$ meson mass which is the natural
scale of the form factors.
As a result, we have
\bea
  &&d\sigma_{12}=-\fr{\alpha^3}{32\pi^4}\, \fr{\rho_2^{++}}
  {s^2 W^2{\bf k}_\bot^2}\, L_2\, {\mathrm{Re}} (F_\pi^*T)
  \fr{d^3p_+d^3p_-}{\varepsilon_+\varepsilon_-}\,,
  \nonumber \\
  &&T=\sum\limits_{ab} \,M_{ab}\,C_1^{ab}\,.
  \label{29}
\eea
The calculation of trace (\ref{22}) which determines $C_1^{ab}$
is given in Appendix~\ref{app:B}.

To present the contribution (\ref{29}) in a compact form, it is
useful to introduce the auxiliary vector ${\bf r}_\perp$ and the
angle $\phi$ between this vector and the vector  ${\bf k}_\perp$
via\fn{The angle $\phi$ is also the azimuthal angle between the
vectors ${\bf p}_+$ and $(-{\bf p}_1)$ in the $\gamma^* \gamma$
c.m.s.}
\bea
 && {\bf r}_\perp = \fr{1}{ 2} \, ({\bf \Delta}_\perp -
  \xi {\bf k}_\perp) = \fr{x_-}{ x} {\bf
  p}_{+\perp} - \fr{x_+}{ x} {\bf p}_{-\perp}\,,
  \nonumber \\
  && {\bf r}_\perp
  {\bf k}_\perp = |{\bf r}_\perp|\, |{\bf k}_\perp |
  \,\cos{\phi}
  \label{31}
\eea
and use the dimensionless quantities
\be
  z_r  =\fr{|{\bf r}_\perp |}{\mu}\,,\quad
  z_k =\fr{|{\bf k}_\perp |}{ \mu}\,,\quad
  d=1-x+\fr{(1-\xi^2) z_k^2}{4(1+z_r^2)}\,.
  \label{32}
\ee
In these notations
\bea
  &&  W^2= 4 \mu^2 \fr{1+z_r^2}{1-\xi^2}\,,\quad
  q_1^2=- \mu^2 \fr{z_k^2}{1-x}\,,
  \nonumber \\
  &&  y_2=\fr{4 \mu^2}{s}\fr{(1+z_r^2) d }{ x (1-x) (1-\xi^2)}\,,
  \label{notations}
  \\
  &&\fr{d^3 p_+d^3 p_+}{\varepsilon_+\varepsilon_-}
  = 4 \pi \mu^4 z_k dz_k z_r
  dz_r d\phi \,\fr{dx}{x} \,\fr{d\xi}{1-\xi^2}\,.
  \nonumber
\eea
We obtain the following expression for the interference
contribution $d\sigma_{12}$:
\bea
  d\sigma_{12} &=&
  \Big[G_{++}\,{\rm Re}\left(F^*_\pi M_{++}\right)
  + G_{+-}\,{\rm Re}\left(F^*_\pi M_{+-}\right) +
  \nonumber \\
  && + G_{0+}\,{\rm Re}\left(F^*_\pi
  M_{0+}\right) \Big]\,\fr{d^3p_+d^3p_-}{\vep_+\vep_-}\,,
  \nonumber \\
  &&G_{ab}=-\fr{\alpha^3}{8\pi^4}\, \fr{\rho_2^{++}L_2}{s^2W^2xz_kd}\,
  g^{ab}\,,
  \nonumber \\
  &&g^{ab}= \sum\limits_{n=0}^3 \, g_n^{ab} \cos{(n \phi)}\,,
  \label{7}
  \\
  &&\rho^{++}_2=\fr{2}{y_2^2}\left(1- y_2 + \fr{1}{2} y_2^2
  \right)\,,
  \nonumber \\
  &&L_2= \ln\fr{|q^2_2|_{\max}(1-y_2)}{m_e^2y_2^2}\,,
  \nonumber
\eea
The nonzero coefficients $g_n^{ab}$ are
\bea
  g_0^{++}&=&(2-x) \xi z_k\,,
  \nonumber \\
  g_1^{++}&=& (2-x)^2 z_r - \fr{2-2x+x^2}{1-x} z_r d\,,
  \nonumber \\
  g_1^{+-}&=&-(2- 2 x+x^2) z_r\,,
  \nonumber \\
  g_2^{+-}&=&- (2-x) \xi  z_k  \,,
  \label{35}
  \\
  g_3^{+-}&=& 2(d-1+x) z_r\,,
  \nonumber\\
  g_0^{0+}&=&  z_r(2-x)\sqrt{2(1-x)}   \,,
  \nonumber \\
  g_1^{0+}&=&- 2 \xi z_k \sqrt{2(1-x)}  \,,
  \nonumber \\
  g_2^{0+}&=& - \fr{2(2-x)}{\sqrt{2(1-x)}}(d-1+x) z_r\,.
  \nonumber
\eea
Let us briefly discuss that result.

We note that
$$
  \fr{\rho_2^{++}}{s^2} \propto \fr{1}{(sy_2)^2} =
  \fr{x(1-x)}{\left[W^2(1-x)+ {\bf k}_\perp^2\right]^2}\,.
$$
Therefore, the effect under discussion does not decrease with growing
$s$, as one could imagine from a first look at Eq.~(\ref{7}).

The coefficient $g^{0+}$ is of the same order as $g^{++}$ and
$g^{+-}$. Near the mass shell the amplitude $M_{0+}\propto
\sqrt{-q_1^2}\sim k_\perp$ [see Eqs.~(\ref{24}),(\ref{26})].
However, in the main region for the charge asymmetry the total
transverse momentum of pion system $k_\perp$ is not small.
Therefore the contribution of the amplitude $M_{0+}$ to the
interference is roughly of the same order of magnitude as the contributions
of the other amplitudes.

The $\pi^+\leftrightarrow\pi^-$ exchange is realized by $\xi \to
-\xi$ and $\phi\to\pi- \phi$ replacements. As it was
discussed above, the contribution $d\sigma_{12}/(d^3p_+ d^3p_-)$
is C--odd, i.e. it
changes its sign under this exchange. Indeed, the coefficients
$g^{++}$ and $g^{+-}$ alter their signs while $g^{0+}$ remains
unchanged. On the other hand, the amplitudes $M_{++}$ and
$M_{+-}$ are unchanged (C--even) while  $M_{0+}$ changes its sign
(C--odd) [see Eq.~(\ref{C2prop})].

According to Eqs.~(\ref{7}),(\ref{35}) the contribution of
amplitude $M_{+-}$ disappears after averaging over the azimuthal
angle $\phi$ (this fact is explained in Appendix~\ref{app:C}):
\bea
  \langle \,d\sigma_{12}\, \rangle_{\phi}&\propto&
  - g_0^{++}\, {\rm Re}\left(F^*_\pi M_{++}\right) -
  g_0^{0+}\,{\rm Re}\left(F^*_\pi M_{0+}\right)
  \nonumber \\
  &=& - (2-x) \times
  \label{37}
  \\
  \Bigl[\xi z_k \Bigr. &{\rm Re}& \Bigl. \left(F^*_\pi M_{++}\right) +
  z_r\sqrt{2(1-x)}\,{\rm Re}\left(F^*_\pi
  M_{0+}\right)\Bigr]\,.
  \nonumber
\eea

For small transverse momentum of the produced pair
$k_\perp\to 0$ our result for $d\sigma_{12}$ coincides with that
of Ref.~\cite{CS} (see Appendix~\ref{app:C}). In this region the asymmetry
in $\xi$ is negligible small compared with that in $v$.

We have obtained our equations in the dominant region of the
effect $k_\perp^2 \sim -q_1^2 \lsim W^2$.
However, our sum $\sum g^{ab}{\rm Re} \left(F^*_\pi M_{ab}\right)$
entering Eq.~(\ref{7}) coincides with the corresponding expression
from Ref.~\cite{Diehl} despite the fact that the latter was
obtained in the quite different kinematical region (at $-q_1^2 \gg
W^2$).

{\bf\bm The term $d\sigma_{13}$} is obtained from the
presented formulae using the substitution rules (see Appendix~\ref{app:C}
for details)
\bea
  \fr{d\sigma_{13}}{ d^3p_+ \,d^3p_-}& = &
  \label{23}
  \\
  & -&
  \fr{d\sigma_{12}}{ d^3p_+ \,d^3p_-} (p_1 \leftrightarrow
  p_2\,,\; p_1^\prime \leftrightarrow p_2^\prime\,,\;q_1
  \leftrightarrow q_2\,)\,,
  \nonumber
\eea
in particular
$$ x_\pm \to y_\pm\,, \quad
  \xi\to\eta \,,
$$
$$
  M_{ab}(q_1,q_2,\Delta)\to (-1)^{a+b}
  \,M_{ba}(q_1,q_2,\Delta)\,.
$$

\section{Two--photon and bremsstrahlung background}
\label{sec:bg}

Let us present now the necessary formulae for the two--photon
$d\sigma_1$ and bremsstrahlung $d\sigma_2$ contributions to the
pion pair production which are background for the considered
asymmetry.

The differential cross section of the {\bf two--photon
contribution} can be written via the helicity amplitudes $M_{ab}$
(\ref{16}) as
\be
  d\sigma_{C=+1} =  \fr{(4\pi \alpha)^2 }{ q_1^2 q_2^2}\,
  \sum\limits_{abcd=\pm 1, \,0}\, M^*_{cd} \,
  M_{ab}\, \varrho_1^{ac}\,\varrho_2^{bd} \fr{d\Gamma}{ 2s} \,.
  \label{19a}
\ee
Here $\varrho_2^{bd}$ is defined in Eq.~(\ref{21}) and
$\varrho_1^{ac}$ is given by a similar expression with the
evident changes $ p_2 \to p_1$, $p_2' \to p_1'$, $ e_2 \to e_1$.

The further calculations repeat those in Sect.~\ref{sec:pi}.  At
fixed value of ${\bf k}_\perp$ there are two regions $(A)$ and
$(B)$ where either $q_2^2 \approx 0$ or $q_1^2 \approx 0$.
Those regions give the dominant contributions in  logarithmic
approximation
\be
  d\sigma_{C=+1}  = \fr{\alpha^2}{32 \pi^5 \,{\bf k}_\perp^2}\,
  \left( T_A +T_B \right)\,
  \fr{d^3 p_+ \,d^3 p_-}{\varepsilon_+\, \varepsilon_-} \,.
  \label{gg}
\ee
In region $(A)$  we can use Eqs.~(\ref{26})--(\ref{qmax}) which
leads to
\be
  T_A =\,
  \fr{\rho_2^{++}}{(sx)^2}\, L_2\,
  \sum\limits_{n=0}^2 T_n\cos(n\phi)\,.
  \label{ggta}
\ee
The coefficients $T_n$ are of the form
\bea
  T_0 &=& \fr{1}{2}\left(|M_{++}|^2 + |M_{+-}|^2  \right) \,
  \left(1-x+\fr{1}{2}x^2\right) +
  \nonumber \\
    &&+|M_{0+}|^2 (1-x)\,,
  \label{gg3}
  \\
  T_1 &=&-{\rm Re}\left[\left( M_{+-}^* -
  M_{++}^*\right) M_{0+}\right]\,\left(2-x
  \right)\,\sqrt{\fr{1-x}{2}}\,,
  \nonumber \\
  T_2 &=&- {\rm Re}\left( M_{+-}^* M_{++}\right)\, (1-x)\,.
  \nonumber
\eea
The helicity amplitudes $M_{ab}$ have be taken in the limit
(\ref{26}).

The contribution $T_B$ of region $(B)$ can be obtained from
$T_A$  substituting $p_1 \leftrightarrow p_2\,,\; p_1^\prime
\leftrightarrow p_2^\prime\, ,\;q_1 \leftrightarrow q_2$
(similar to Eq.~(\ref{23}) but without changing  sign).

A detailed analysis of these equations~\cite{BGMS}  shows that
the pairs produced via the two--photon mechanism are
concentrated at small values of $k_\perp$
\be
  d\sigma_{C=+1}\propto \fr{d{\bf k}_\perp^2}{{\bf k}_\perp^2}
  \ln\fr{{\bf k}_\perp^2s}{m_e^2W^2}
  \label{bckg2qualk}
\ee
and they are distributed almost uniformly in  rapidity
\be
  d\sigma_{C=+1}\propto
  \fr{dx}{x}\,\fr{dy}{y}\,.
  \label{bckg2qualx}
\ee

For the {\bf bremsstrahlung contribution} $d\sigma_{C=-1}= d\sigma_2 +
d\sigma_3$ we use the results of Ref~\cite{KS}. For pions flying
along electrons, the dominant contribution is given by the
amplitude ${\cal M}_2$ taken in the limit (\ref{26}). In
logarithmic approximation we have
\bea
  d\sigma_{2} &=&  \left|{\cal
  M}_2 \right|^2 \fr{d\Gamma}{2s}=
  \fr{\alpha^4}{8 \pi^3} \,|F_\pi|^2 \times
  \label{br}
  \\
  &&
  \fr{x^2\,\rho_2^{++}}{W^2 [W^2 (1-x) +{\bf k}_\perp^2+m_e^2 x^2]^2}\, L\,
  T_-\, \fr{d^3 p_+\,d^3 p_-}{\varepsilon_+\, \varepsilon_-}\,,
  \nonumber\\
  T_-&=& \left(1- \fr{4\mu^2}{W^2} \right)\, \left[y^2 +\left(y_2-
  \fr{W^2}{s}\right)^2\right] -
  \nonumber \\
  && -y^2 \eta^2 - (y\eta+y_2x\xi)^2
  \nonumber \\
  L&=& \ln{\fr{s^2\,(1-x)(1-y_2)}{m_e^2\,[W^2+sy_2(1-x)]}}\,.
  \nonumber
\eea
The denominator $[W^2 (1-x) +{\bf k}_\perp^2+ m_e^2 x^2]^2$
shows that this contribution is dominated by small ${\bf
k}_\perp^2$ and  large  $x$ (i.e. $1-x \ll 1$).

For  pions flying along positrons  the corresponding contribution
$d\sigma_{3}$ is obtained via $d\sigma_{3}= d\sigma_2 (x
\leftrightarrow y, \, y_2 \leftrightarrow x_1, \, \xi
\leftrightarrow \eta)$.

\section{The $\gamma^* \gamma \to \pi^+ \pi^-$ subprocess }
\label{sec:ggpipi}

For the pion pair production the strong interaction effects are of
primary interest.  Nevertheless, the pure {\bf Born QED model}
(point--like pions) gives a reasonable description of the
squared two--photon amplitude at $W\lsim 1$~GeV. On the contrary,
the QED amplitude itself is real whereas the phase shifts of the correct
$\gamma\gamma\to \pi^+\pi^-$ amplitudes coincide with those of
elastic $\pi\pi$ scattering (at least, at $W<520$ MeV).
The QED amplitudes $M_{ab}$ entering
Eq.~(\ref{29}) are (see Appendix~\ref{app:A})
\bea
  M_{++}^{QED} &=& 8 \pi \alpha  -M_{+-}^{QED}\,,
  \nonumber \\
  M_{+-}^{QED} &=& 8 \pi \alpha \fr{(1-x) z_r^2}{ (1+z_r^2) d}\,,
  \label{39}
  \\
  M_{0+}^{QED} &=& 4 \pi \alpha\xi \fr{ \sqrt{-2
  q_1^2}}{\mu}\, \fr{(1-x)(d- 2 + 2 x ) z_r} {(1+z_r^2)d^2}\,,
  \nonumber \\
  F_\pi^{QED}&=&1\,.
  \nonumber
\eea
In the next Section we obtain numerical results for that model.
It allows us to develop a better understanding of the potentiality
of future experiments.

With increasing dipion effective mass $W$ above the threshold the
strong interaction effects become more essential. At $W\gsim 1$
GeV these effects dominate and the main contribution is given by
resonances. The pion form factor entering the bremsstrahlung
amplitude is experimentally well studied  using the reaction
$e^+e^-\to\pi\pi$. Using the charge asymmetry the main task in
this domain is to study the resonances in the two--photon channel,
i.e. the different $f_0$'s and $f_2$. The nature of $f_0$
resonances is a subject of discussions till now, and different
models of the resonance origin can lead to a different
$W$--dependence of their phases. Those models can be tested using
the discussed asymmetry. For the $f_2$ resonance the value of its
amplitude for the two--photon production with total helicity 0 can
be obtained studying the longitudinal charge asymmetry (which
practically does not depend on the amplitude with total initial
helicity 0).

Near the resonances some of amplitudes $M_{ab}$ are enhanced
compared to their QED values. Besides, we have for the pion form
factor $|F_\pi(W^2)| > 1$ in a wide enough region of interest. A
detailed study of the effect with a realistic $\ggam\to
\pi^+\pi^-$ amplitude and a discussion of the potentiality of such
experiments to study the phase shifts of $\pi\pi$ scattering and
the resonance nature will be presented elsewhere.

\section{Analysis of results}
\label{sec:analysis}

\subsection{Studied quantities}

As it was discussed before, the contribution $d\sigma_{12}$
dominates for pion pairs flying along the initial electron, i.e.
at large values of $x$. On the other hand, $d\sigma_{13}$
dominates for dipions flying along the initial positron, i.e. at
large values of $y$. Since  $xy=(W^2+{\bf k}_\bot^2)/s$, we can
introduce the characteristic value
\be
  x_0=\sqrt{\frac{W^2+{\bf k}_\perp^2}{s}}
  \label{x0}
\ee
and find that the total longitudinal momentum of the pion pair
in the c.m.s. of the colliding electrons and positrons
$k(p_2-p_1)/\sqrt{s} \propto x-y$ is positive at $x>x_0$ and
negative at $x<x_0$. Therefore, $|d\sigma_{12}| \gg
|d\sigma_{13}|$ at $x\gg x_0$ and $|d\sigma_{13}| \gg
|d\sigma_{12}|$ at $x\ll x_0$.

In the region $x\sim y \sim x_0$ both contributions are of the
same order, $|d\sigma_{12}| \sim |d\sigma_{13}|$, but their
distributions over the longitudinal variable $K_-$ and the
transverse variable $v$ have different properties. Indeed, $K_-$
is antisymmetric whereas $v$ is symmetric under $p_1
\leftrightarrow p_2$ exchange. Having in mind relation (\ref{23}),
we conclude that
$$
  \fr{d\sigma_{13}}{dK_-}=\fr{d\sigma_{12}}{dK_-}
  (p_1 \leftrightarrow p_2\,,\;
   p_1^\prime \leftrightarrow p_2^\prime\,,\;q_1 \leftrightarrow q_2) \,,
$$
$$
  \fr{d\sigma_{13}}{dv} = - \fr{d\sigma_{12}}{dv} (p_1 \leftrightarrow p_2\,,\;
  p_1^\prime \leftrightarrow p_2^\prime\,,\;q_1 \leftrightarrow q_2)\,.
$$

To take into account the described properties and to summarize
different contributions, it is natural to introduce the following
quantities related to the charge asymmetry of pions\footnote{Below
we use the standard step functions $\theta(x)$ and $\epsilon(x) =
\theta(x) - \theta(-x)$. For example, $\Delta\sigma_K$ is the
difference between cross sections for events with $p_{+z}>p_{-z}$
and that with $p_{+z}<p_{-z}$ in the $e^+e^-$ c.m.s.}
\be
  \Delta\sigma_K=\int\limits_{\cal D}\epsilon(K_-)d\sigma\,, \quad
  \Delta\sigma_v=\int\limits_{\cal D}\epsilon(v)\epsilon(x-y)d\sigma\,.
  \label{sigw}
\ee
In these definitions we denote by ${\cal D}$ the kinematical
region in phase space given  by the detector array and suitably
chosen cuts (certainly, it is necessary to test that this region
is symmetric in $K_-$ and $v$).

The background for these effects is found integrating the
two--photon and brems\-strah\-lung contributions over the same
region $\cal D$:
\be
  \Delta\sigma_B=\int\limits_{\cal D}
  \left( d\sigma_{C=+1}+ d\sigma_{C=-1}\right).
  \label{background}
\ee

\subsection{Numerical analysis}

Below we consider the charge asymmetry effects within
the QED model of point--like pions.
First, we present in Table~\ref{tab:1}
\begin{table}[!htb]
  \begin{center}
    \caption{Pion charge asymmetry signals and background at
             different c.m. energies}
    \label{tab:1}
    \begin{tabular}{ccccc}\hline
      $\sqrt{s}      $, GeV&1   &    4 & 10  &200    \\\hline
      $\Delta\sigma_K$, pb &-6.1&  -26 &-35  &-56    \\
      $\Delta\sigma_v$, pb &3.3 &   17 & 27  & 51    \\
      $\Delta\sigma_B$, pb &420 & 2900 &6200 & 27000 \\\hline
    \end{tabular}
  \end{center}
\end{table}
the integrated pion charge
charge asymmetry in the variables $K_-$ and $v$ (signals $S$)
and the background of two--photon and bremsstrahlung production
(background $B$) at different c.m. energies $\sqrt{s}$.

\begin{figure}[!htb]
  \centerline{\epsfig{file=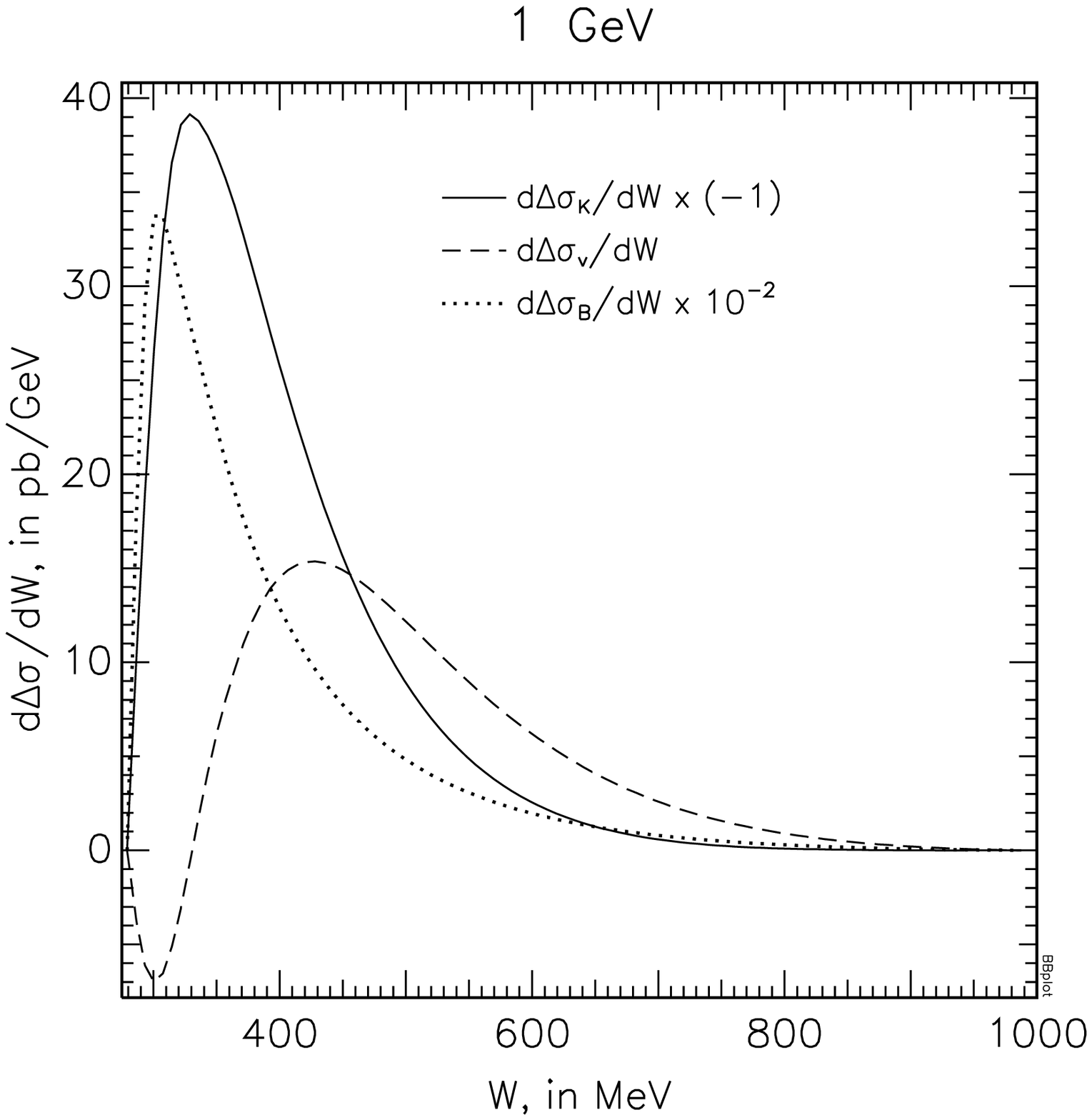,width=6cm}}
  \centerline{\epsfig{file=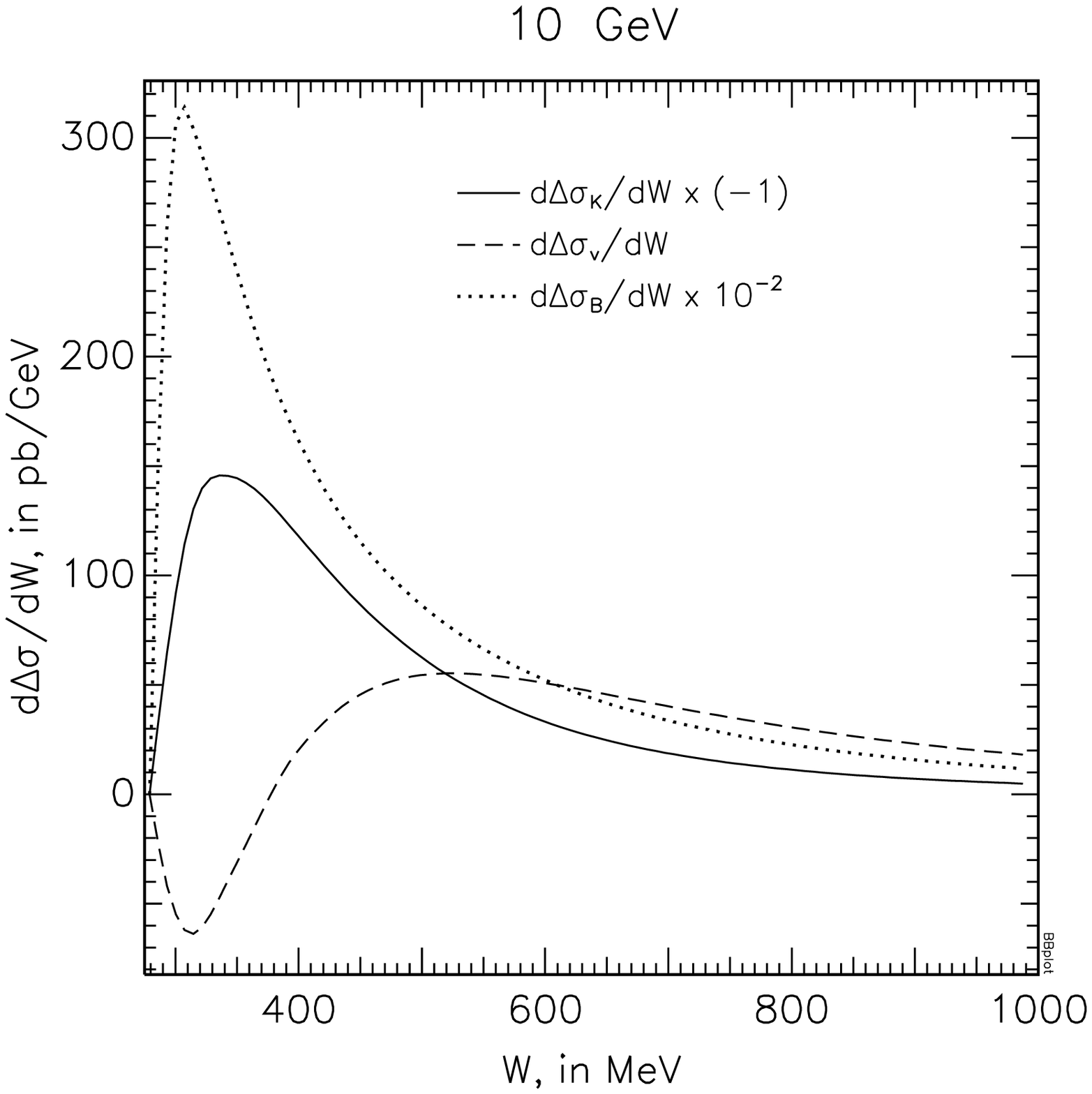,width=6cm}}
  \caption{
           Contributions $\Delta \sigma_K$ and $\Delta \sigma_v$
           (\protect\ref{sigw}) and background at
           $\protect\sqrt s = 1$ and 10 GeV vs. $W$.
          }
  \label{fig:4}
\end{figure}

Using the charge asymmetry one can study the two--photon amplitude
and its phase shift as function of  the effective mass of the
$\pi^+ \pi^-$ system $W= \sqrt{k^2}$. To see the potentiality of
such a study, we present in Fig.~\ref{fig:4}
the distribution of the signal and the background over $W$ for two
collider energies. Both signal and background are concentrated
near the threshold where the longitudinal asymmetry $|d\Delta
\sigma_K /dW|$ is considerably larger than the transverse one
$|d\Delta \sigma_v /dW|$. At $W>400\div 500$ MeV (depending on
$\sqrt{s}$) the transverse asymmetry dominates over the
longitudinal one. Nevertheless, having in mind the results for the
muon charge asymmetries presented below we discuss in following
the longitudinal asymmetry mainly.

We consider two typical intervals of effective mass values
(over which we integrate):\\
1. $W=300 \div 350$ MeV --- near the threshold
where QED is approximately valid,\\
2. $W=475 \div 525$ MeV --- far from the threshold and resonances
where one hopes to describe the modules of the two--photon
amplitudes reasonably within QED whereas the bremsstrahlung
amplitude is enhanced compared to its QED value due to the $\rho$ meson
resonance. For this region we expect that our numbers
underestimate the effect.

The signal/background ratio (S/B) is introduced as
\be
  \fr{S}{B}=\fr{|\Delta\sigma_S|}{\Delta\sigma_B}\quad
  \mbox{ with } S=K_{-},\;v.
  \label{SB}
\ee
Besides, it is useful to consider the statistical
significance (SS) of the effect.  This quantity is expressed
via the number of events for the effect $N_S= {\cal
L}|\Delta\sigma_S|$ and background $N_B={\cal L}\Delta\sigma_B$
as
\be
  SS = \fr{N_S }{ \sqrt{N_B}}
  \label{SS}
\ee
where $\cal L$ is  the integrated luminosity of the collider.
For the luminosity we use numbers proposed for the
 DA$\Phi$NE and PEP~II colliders.
We now demonstrate that the S/B and SS  quantities can be
considerably improved with suitable cuts on  $k_\perp$ and $x$.

\begin{figure}[!htb]
  \centerline{\epsfig{file=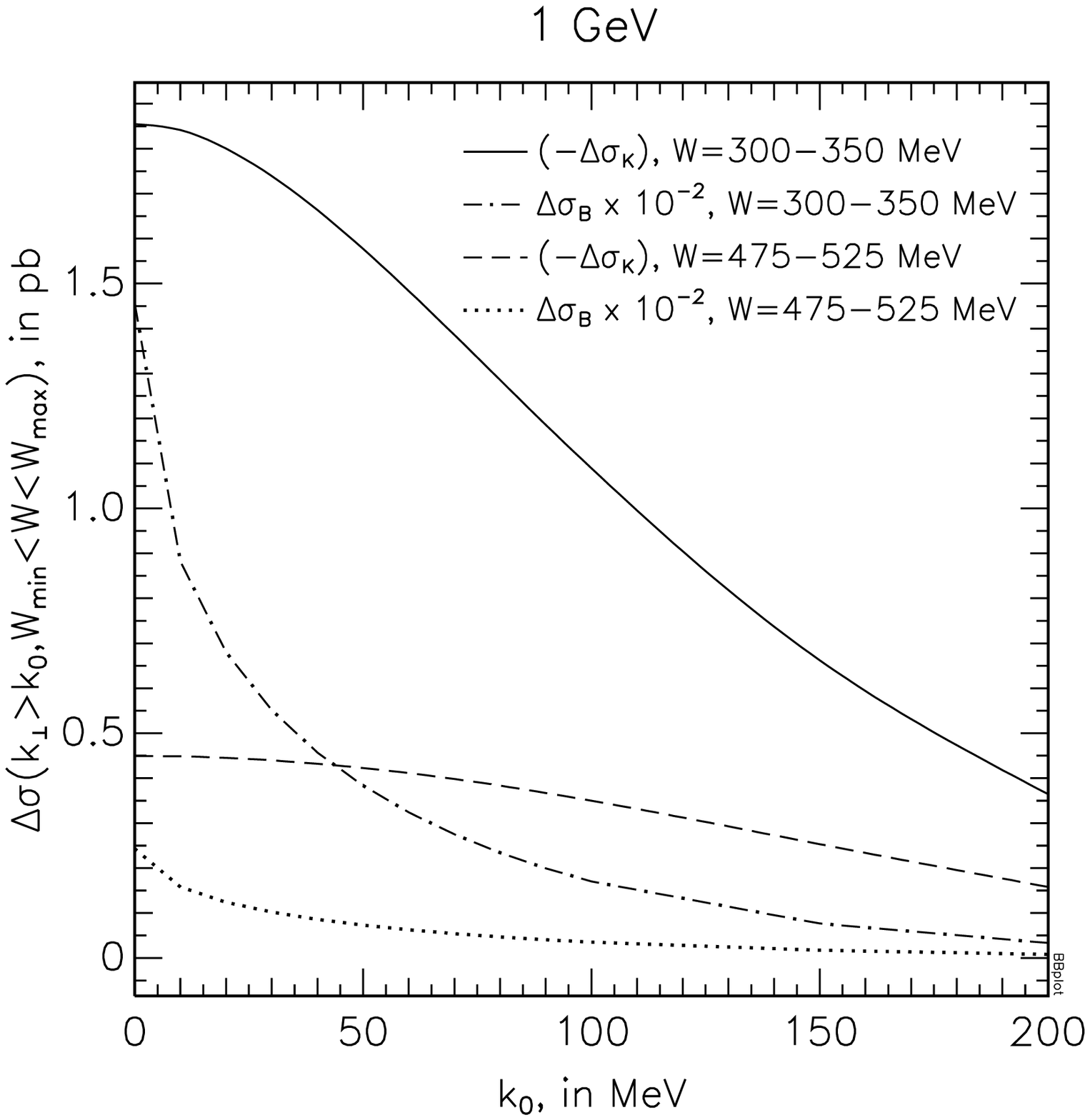,width=6cm}}
  \centerline{\epsfig{file=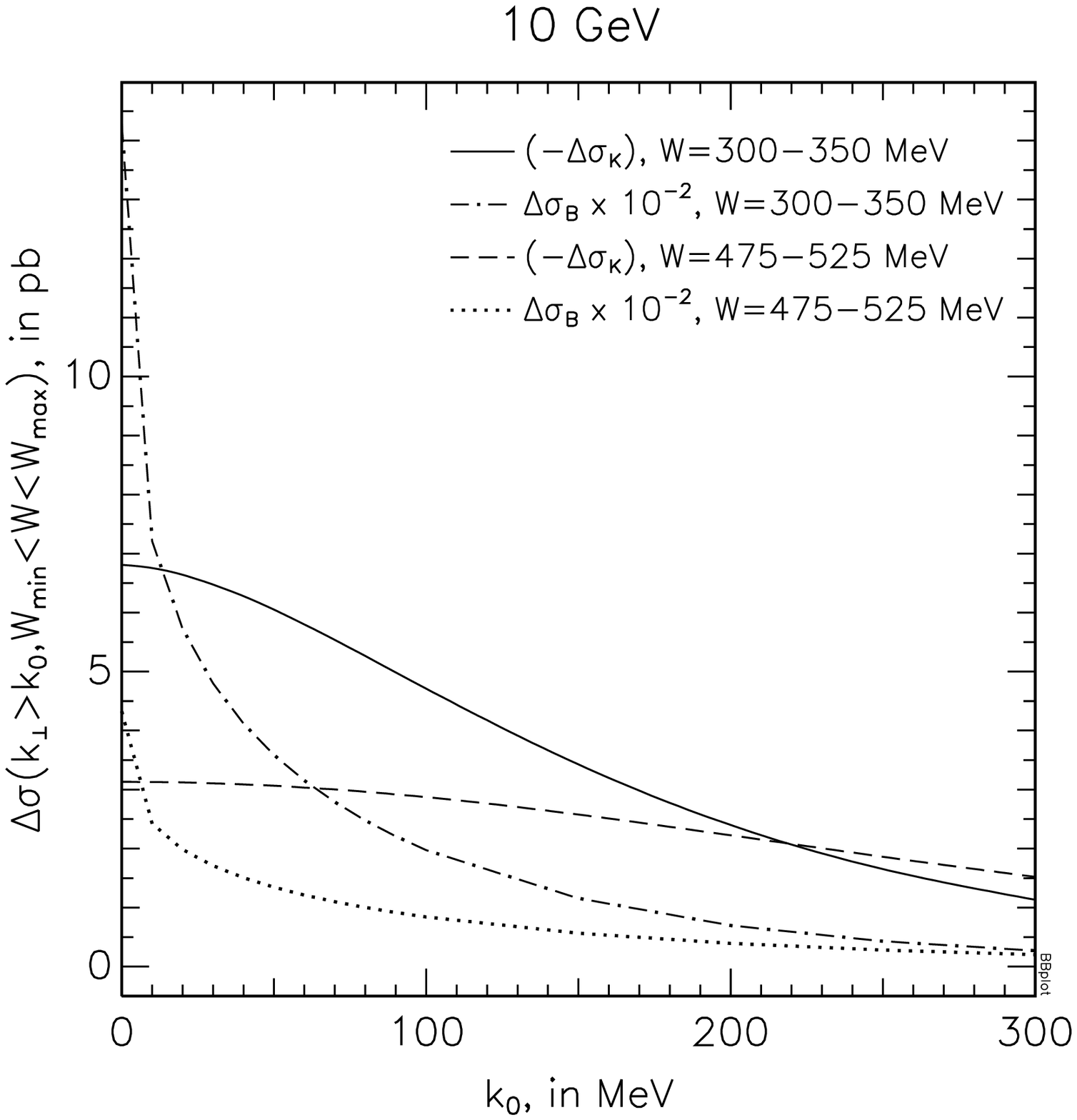,width=6cm}}
  \caption{
           Contributions $\Delta \sigma_K$ and background at
           $\protect\sqrt s  =1$ and 10 GeV for the interval
           $W= 300 \div 350$ MeV and $W=475 \div 525$ MeV
           and $k_\perp > k_0$ vs. $k_0$.
          }
  \label{fig:5}
\end{figure}

The variable ${\bf k}_\perp^2$ describes both the
transverse motion of the dipion and the virtuality of the
photon. The charge asymmetry effect (\ref{7}),(\ref{35})
vanishes at small $|{\bf k}_\bot|$, $d\sigma_{12}\propto d|{\bf
k}_\bot|$. On the contrary, the two--photon contribution is
singular at $|{\bf k}_\bot|\sim 0$, see Eq.~(\ref{bckg2qualk}).
In Fig.~\ref{fig:5}
we present signal and background integrated over  $k_\perp >
k_0$ (with $k_0$ being a cut--off from below). One observes
that with increasing $k_0$ the background drops considerably
faster than the signal. Therefore, some cut at small ${\bf
k}_\bot^2= k_0^2$ is  desirable, the best cut on the pair
transverse momentum $k_0$ depends on $s$ and $W^2$.

\begin{figure}[!htb]
  \centerline{\epsfig{file=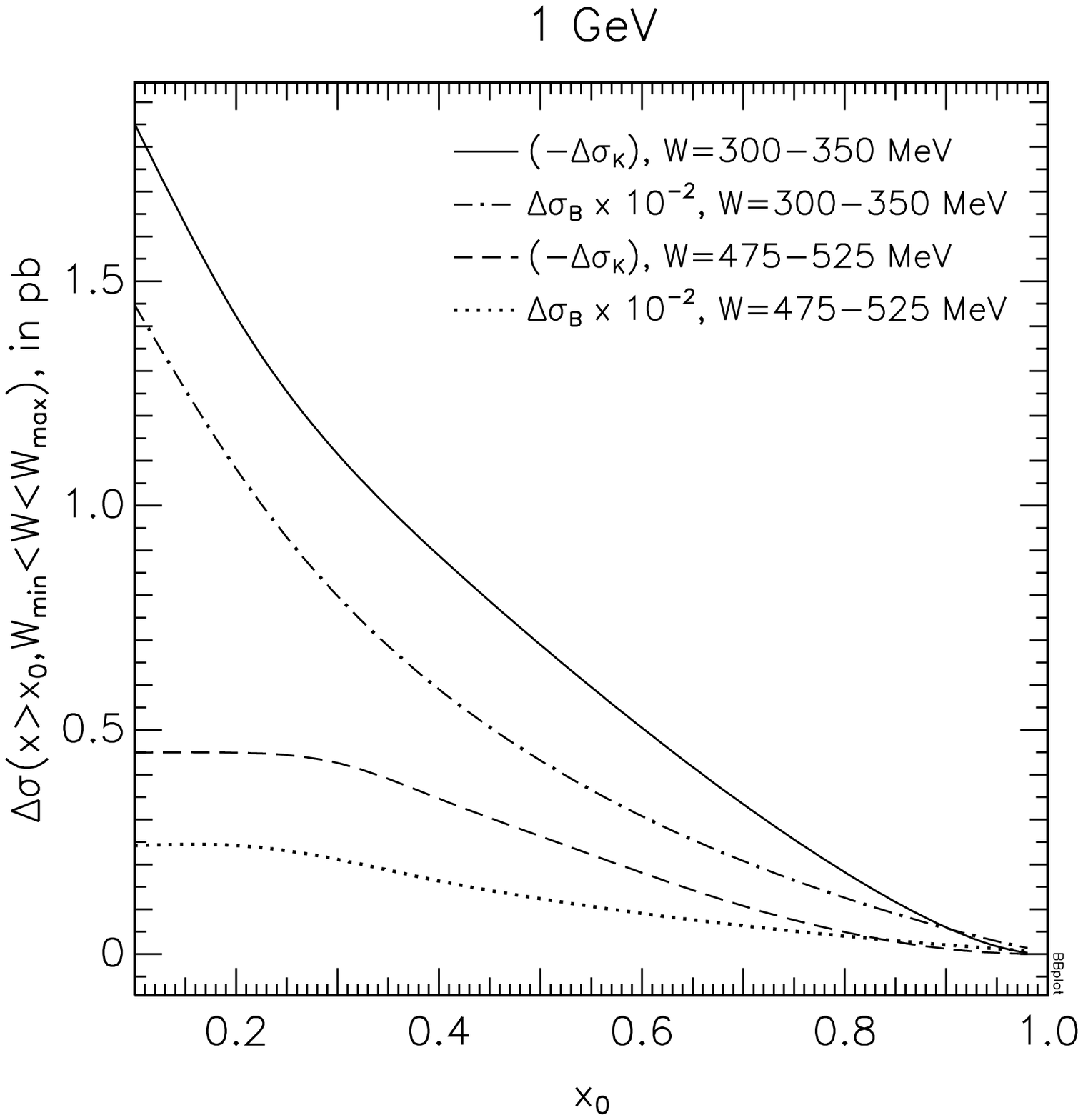,width=6cm}}
  \centerline{\epsfig{file=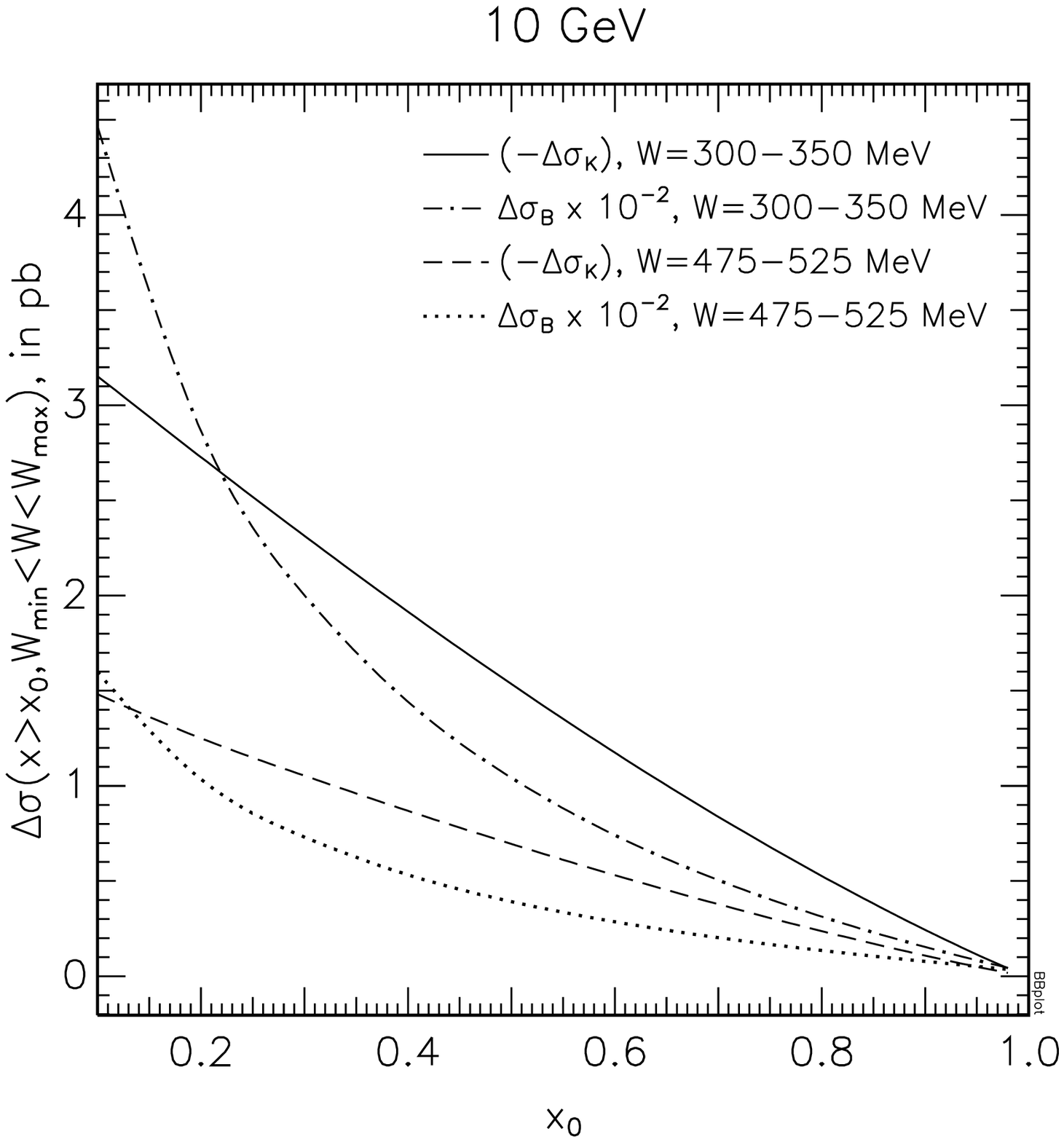,width=6cm}}
  \caption{
           The contributions $\Delta \sigma_K$ and background at
           $\protect\sqrt s  =1$ and 10 GeV for the intervals
           $W= 300 \div 350$ MeV and $W=475 \div 525$ MeV and $x > x_0$
           vs. $x_0$.
          }
  \label{fig:6}
\end{figure}

The variable $x$ describes the dipion motion along the collision
axis. The factor $W^2d \equiv W^2(1-x)+{\bf k}_\perp^2$ in the
denominator of Eq.~(\ref{7}) shows that in the interference the
dipions tend to be concentrated at $x\sim 1$. On the contrary,
in the two--photon production the $x$--distribution of the
dipions is proportional to $1/x$.  Therefore, some cut at not
too low $x$ would be desirable.  On the other hand, the
bremsstrahlung contribution is concentrated near $x=1$ more
strongly than the charge asymmetry contribution.  Moreover, the
values of $x$ very close to 1 contribute only weakly to the
charge asymmetry. Therefore, an additional cut at $x$ near 1 is
suggested. In Fig.~\ref{fig:6}
these features are demonstrated
for signal and background contributions integrated over $x>x_0$.

To show
the effect of both cuts we present in Table~\ref{tab:2}
\begin{table*}[!htb]
  \begin{center}
  \caption{Effect of cuts in $k_\perp$ and $x,\,y$}
  \label{tab:2}
  \begin{tabular}{cccccccc}\hline
  $\sqrt{s}$ & ${\cal L}$ & $W$      & cuts         &$\Delta\sigma_B$&
  $\Delta \sigma_K$ &$S/B$ & $SS$  \\
             &  fb$^{-1}$ &      MeV &              &             pb &
                 pb &   \%  &        \\\hline
  1 GeV      & 5        & 300   &no cuts         &   145                &
        -1.85           &1.3 &  11   \\
  DA$\Phi$NE &          & $\div$   &$k_\bot>100$ MeV,& 14.6                 &
  -1.07                 &7.3 &20      \\
  &&350&$0.4<x,y<0.9$&&&&\\
  \hline
  10 GeV& 30            &475 &no cuts&433&-3.13&0.72&8.2\\
  PEP-II& & $\div$ & $k_\bot>150$ MeV,& 17.2&
  -1.62&9.5&68\\
  &   &  525 &$0.3<x,y<0.95$&&&&\\
  \hline
  \end{tabular}
  \end{center}
\end{table*}
some examples considering cuts in $k_\bot$ and
two symmetrical regions in pion rapidity (contributions of pions
flying along initial electron momentum $x_1>x>x_2$ and
initial positron momentum $x_1>y>x_2$).
For the DA$\Phi$NE collider we consider the region of
small effective mass of dipion $W= 300\div350$ MeV. The used cuts
improve the $S/B$ by a factor about 5 and the $SS$ -- by a factor
about 2. For the PEP-II collider we consider an intermediate
mass region $W= 475\div 525$ MeV. The used cuts improve both the
$S/B$  and the $SS$ by about one order of magnitude. It
is natural to expect that the same type of improvement will take
place at $W\sim 1$ GeV.

\begin{figure}[!htb]
  \centerline{\epsfig{file=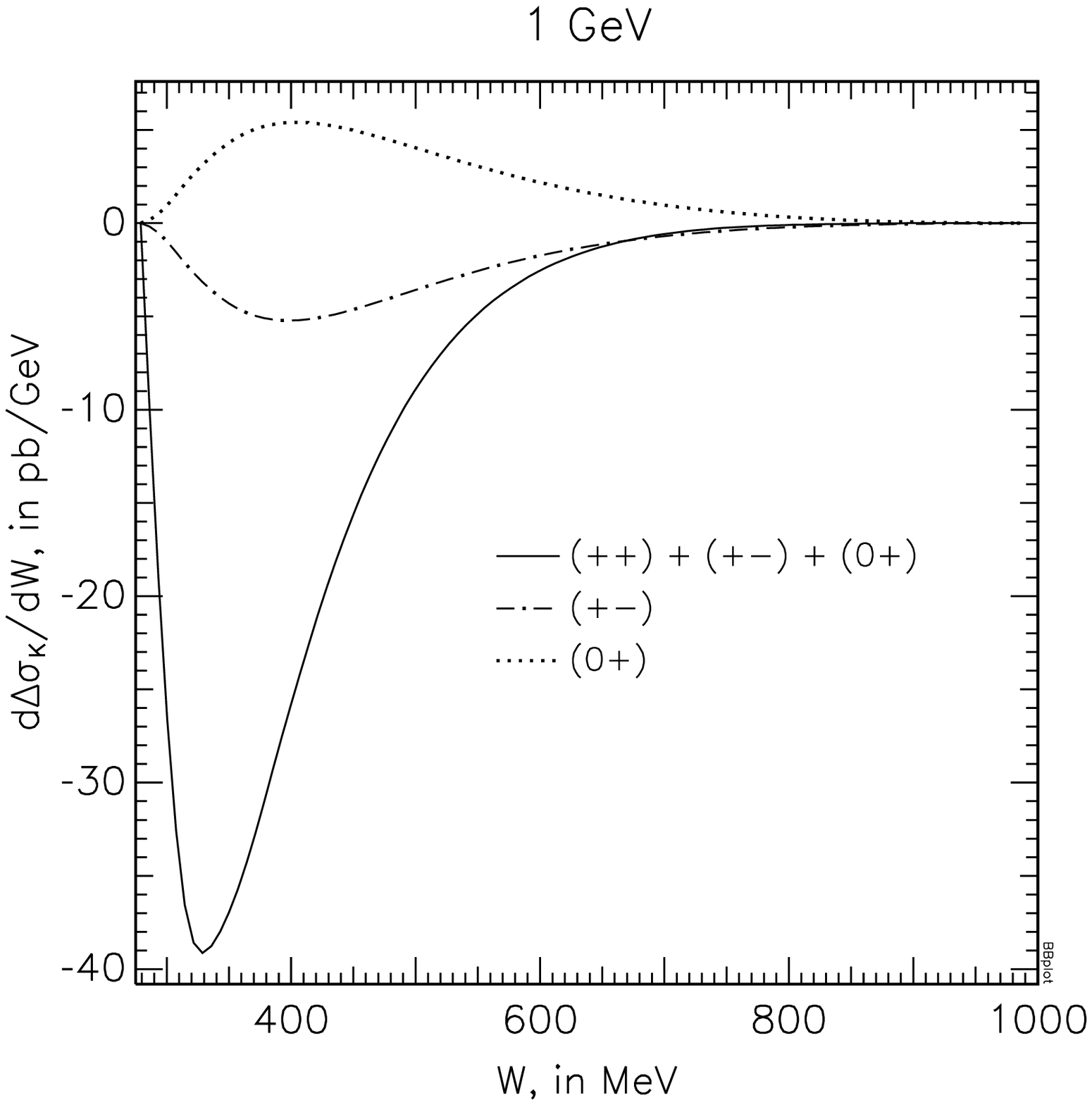,width=6cm}}
  \centerline{\epsfig{file=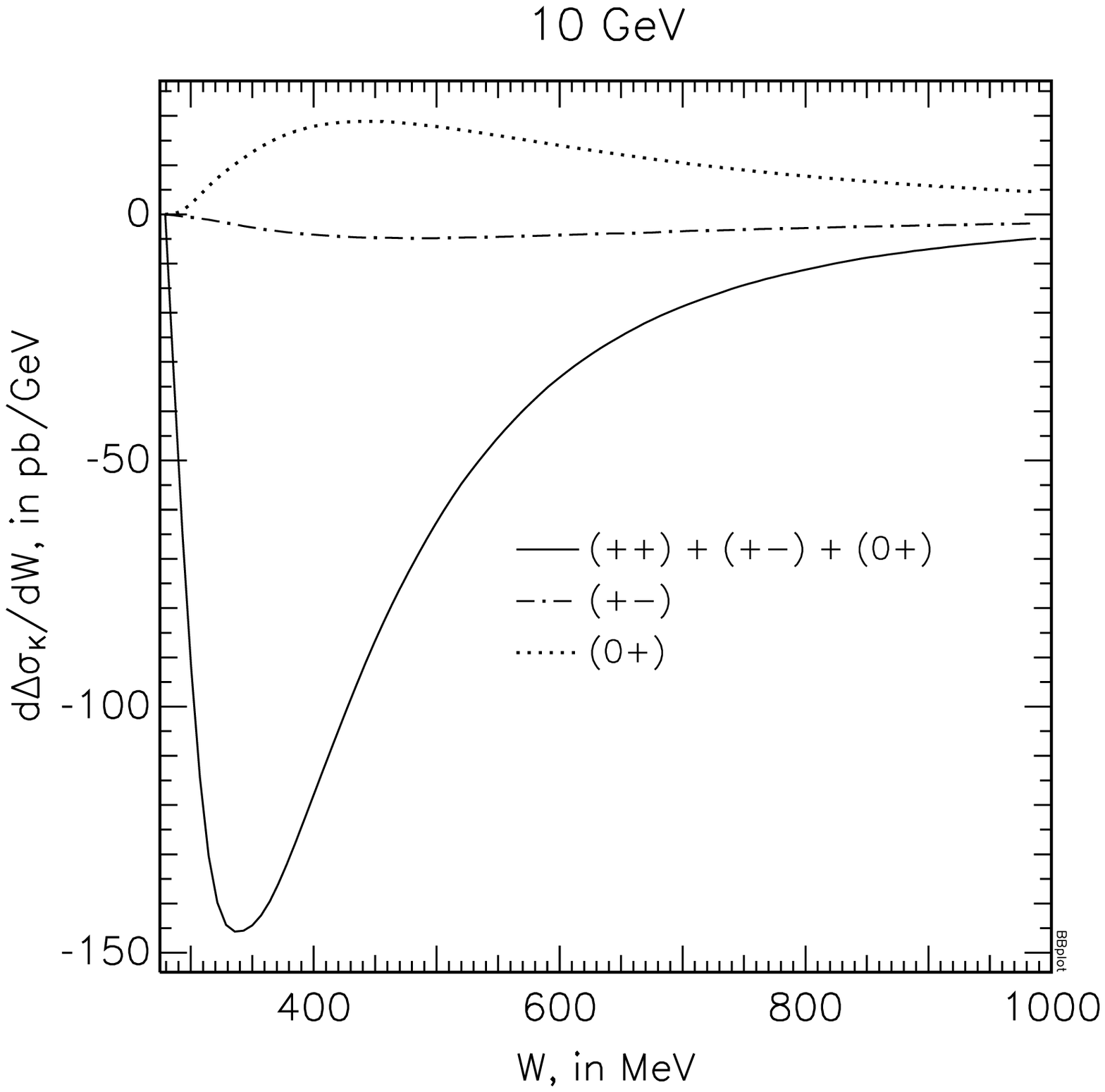,width=6cm}}
  \caption{
            Contributions of the helicity amplitudes $M_{ab}\equiv(ab)$
            to the distribution $d\Delta \sigma_K / dW$
            (\ref{sigw}) vs. $W$
            at $\protect\sqrt s  = 1$ and 10 GeV.
          }
  \label{fig:7}
\end{figure}

For the physical analysis of the results it is useful to consider the
individual contributions of different helicity amplitudes $M_{ab}$
to the charge asymmetry. The results for the longitudinal
asymmetry $d\Delta \sigma_K / dW$ are shown in
Fig.~\ref{fig:7}.
In this distribution the amplitude $M_{++}$ is dominant whereas
$M_{+-}$ contributes only weakly to the asymmetry
(in accordance with the discussion at the end of Sect.~\ref{sec:pi}).
The last contribution can be
even stronger suppressed,  excluding pion pairs with small
longitudinal momentum in the $e^+e^-$ c.m.s. additionally.
Therefore, the distribution over $K_-$ allows us to obtain a clean
information about the amplitude $M_{++}$.

\begin{figure}[!htb]
  \centerline{\epsfig{file=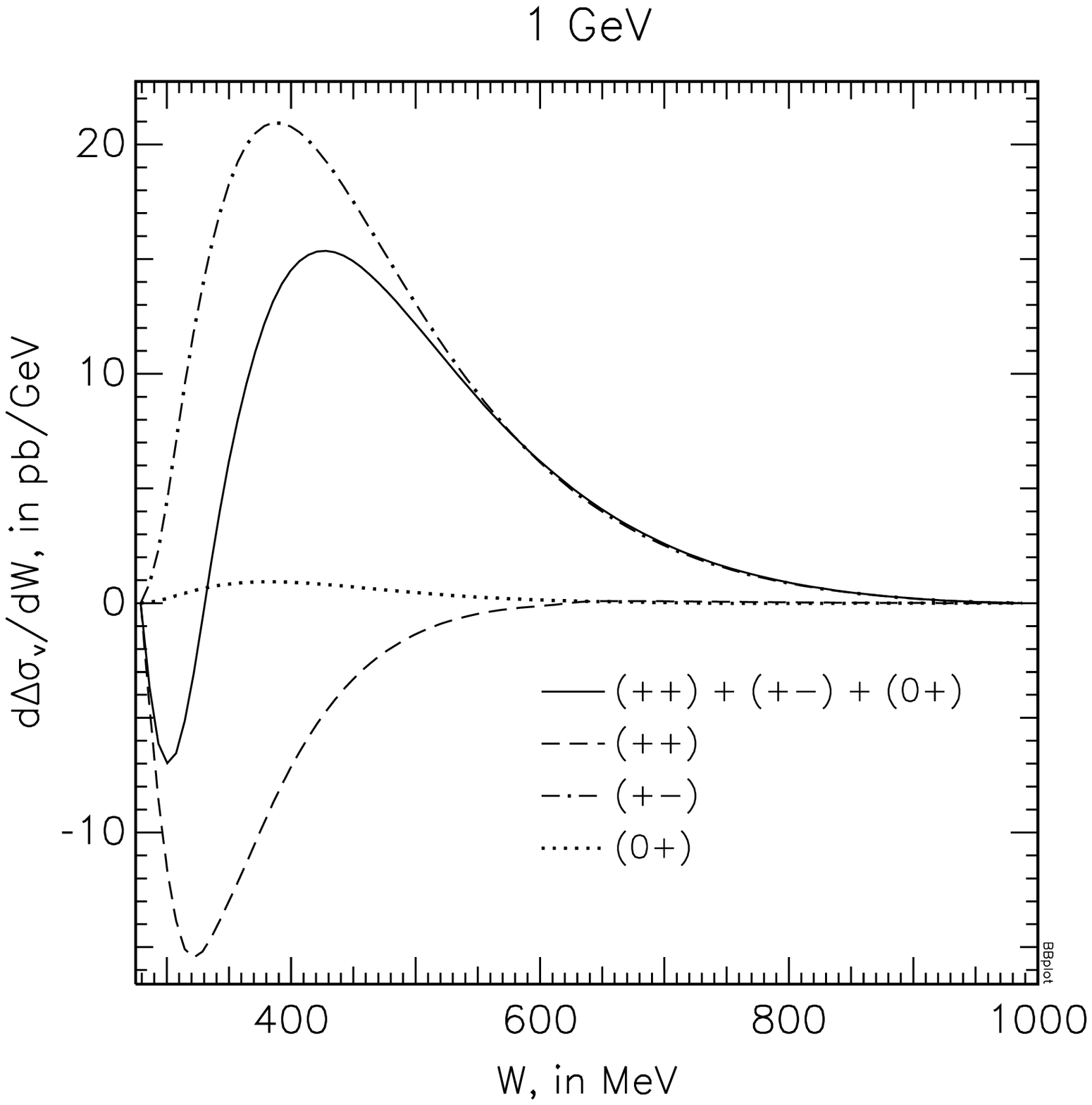,width=6cm}}
  \centerline{\epsfig{file=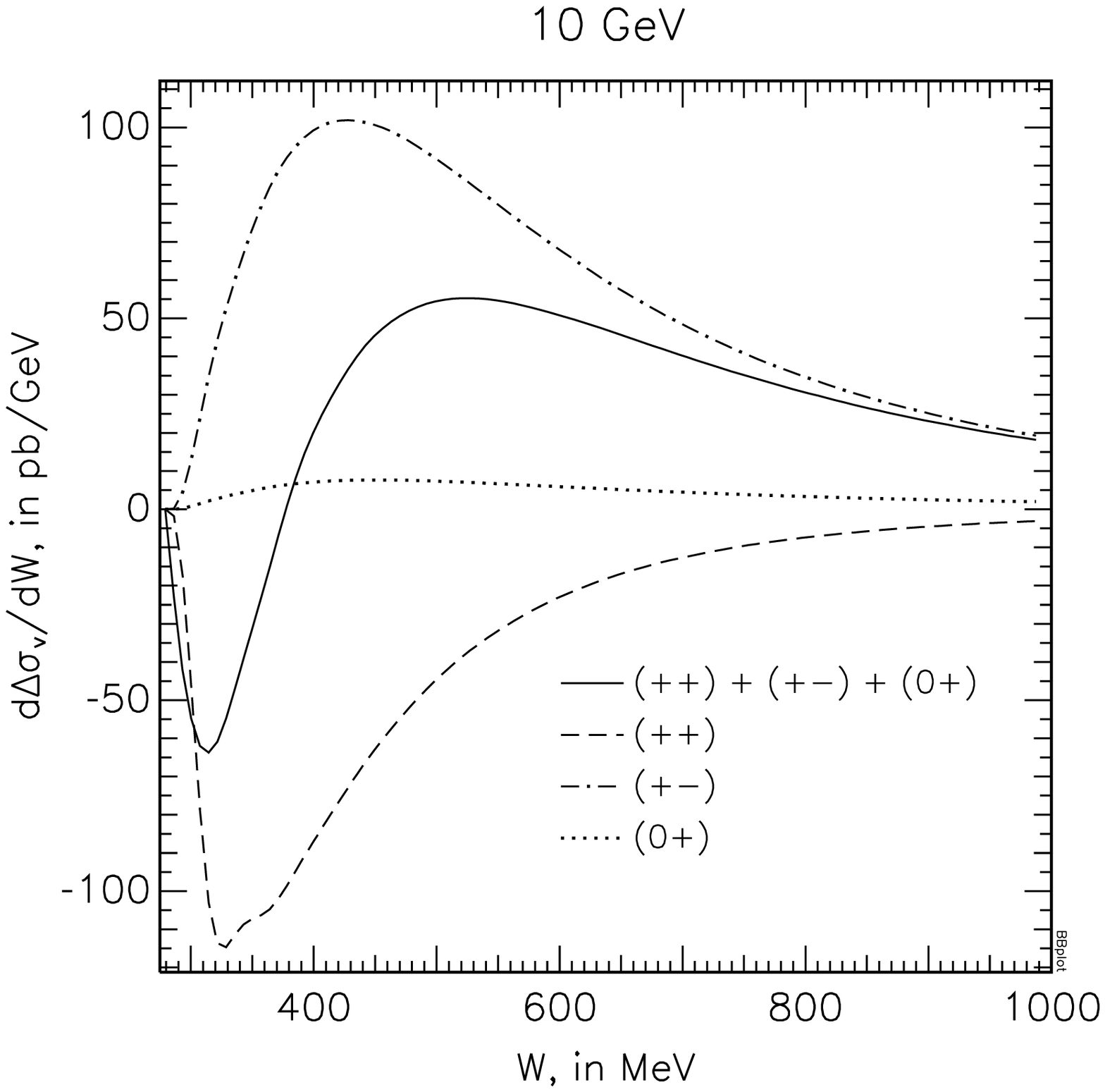,width=6cm}}
  \caption{
           Same as in Fig.~\protect\ref{fig:7} for the
           distribution $d\Delta \sigma_v / dW$ (\protect\ref{sigw})
          }
  \label{fig:8}
\end{figure}

In the transverse distribution $d\Delta \sigma_v/dW$ (Fig.~\ref{fig:8})
the contribution of the $M_{++}$ and $M_{+-}$ amplitudes are of the
same order, partially compensating each other. Therefore, the
combined distributions over $K_-$ and $v$ can give
complementary
knowledge about individual contributions of the $M_{++}$ and
$M_{+-}$ helicity amplitudes.

Finally let us notice, that in the $K_-$--distribution  the
contribution of the amplitude
with one scalar photon $M_{0+}$ is not negligible.

\subsection{Weighted cross sections}

For future studies it is useful to look at the presented
analysis from a more general point of view. The distributions of
produced pions contain a charge asymmetric part. To extract it,
we consider each event with the $C$--odd weight function
$\epsilon(K_-)$ or $\epsilon(v)$. So, our asymmetry
$\Delta\sigma_K$ and $\Delta\sigma_v$ defined in
Eqs.~(\ref{sigw}) can be considered as ``weighted cross
sections'' with these weights. Certainly, to extract the
asymmetric part of the cross section, one can use also other
$C$--odd weight functions. It seems to be attractive to explore
more smooth weight functions, for example, given by factors
$K_-$ and $v$ instead of $\epsilon(K_-)$ and $\epsilon(v)$,
i.e. to introduce weighted cross sections
\be
  \Delta\sigma^C_K=\int\limits_{\cal D}\,K_-\,d\sigma\,, \quad
  \Delta\sigma^C_v=\int\limits_{\cal D}\,v\,\vep(x-y)d\sigma\,.
  \label{sigcw}
\ee
They (or similar quantities) may be more suitable for a
theoretical analysis, generalizations and data processing. In
particular, the signs of small differences
$p_{+z}-p_{-z}$ or ${\bf p}^2_{+\bot} - {\bf
p}^2_{-\bot}$ cannot be reliably established from the data. 
However, the proposed weighted cross sections (\ref{sigcw})
are weakly sensitive to those small values.

The background for these ``cross sections'' is given by the total
weighted cross section of the process in the same kinematical
region,
\bea
  \Delta\sigma^C_{BK}&=& \int\limits_{\cal
  D}\,|K_-|\,\left(d\sigma_{C= +1}+ d\sigma_{C=
  -1}\right)\,,
  \nonumber \\
  \Delta\sigma^C_{Bv}&=& \int\limits_{\cal D}\, |v|\,
  \left(d\sigma_{C= +1}+ d\sigma_{C= -1}\right)\,.
  \label{sigCB}
\eea

\section{Process  $e^- e^+ \to e^-e^+\mu^+\mu^-$}
\label{sec:muon}

The process $e^- e^+ \to e^-e^+\mu^+ \mu^-$ can give an essential
background while studying the the dipion production.

The charge asymmetry of muons in this process
has been studied for the first time at small $k_\bot$ in
Ref.~\cite{CS}. The muon asymmetry without that limitation was
obtained in Ref.~\cite{KLMS} (in the same logarithmic accuracy
as it is used here).
We use these results as given in review~\cite{BFKK} and
transform them to a form convenient for analysis\footnote{Note
two misprints in Ref.~\cite{BFKK}: First, diagram ({\it
b}) in Fig. 4.1 should be replaced by diagram ({\it c}) and vice
versa. Second, in the statement after Eq.~(4.36) ``$d\sigma_{ac}$
may be derived from (4.36) through the substitution (4.4)'' one
has to add ``changing the overall sign'' -- see Appendix~\ref{app:C} for the
pion case.}. In notations (\ref{31}),(\ref{32}) (where $\mu$ is
now the muon mass) we have
\bea
   d \sigma_{12}&=& \fr{ 2\alpha^4}
  { \pi^3} \fr{ \rho_2^{++} L_2} {s^2 W^2 x z_k(1+z_r^2)d^2} \times
  \nonumber \\
  &&\left[  \sum\limits_{n=0}^3\, t_n \cos (n \phi) \right]
  \,
  \fr{d^3p_+d^3p_-}{\varepsilon_+\varepsilon_-} \,,
  \label{38b}
  \\
  t_0 &=& t_2 +  (2-x) \xi \,z_k d\,,
  \nonumber \\
  t_1 &=& \fr{z_r}{4} \Bigg\{ (1-x)  \left[ 8 \,(1-x) + 10\,
   (1-\xi^2) z_k^2\right] +
   \nonumber \\
   &&+(2-2x+x^2) \left[\fr{d-1+x}{1-x} (1+\xi^2)\, z_k^2 + \right.
   \nonumber \\
   && \left.+ 4 \left( - (d-2+2x)   + (1-x) \fr{1+\xi^2}{1-\xi^2}
  (1+z_r^2)\right)\right]  \Bigg\}
  \nonumber \\
  t_2 &=&    (2-x) \xi  (d- 2 + 2x) z_k z_r^2\,,
  \nonumber \\
  t_3 &=& 2\, (1-x) (d - 1 + x ) z_r^3\,,
  \nonumber
\eea
with $L_2$ given by Eq.~(\ref{7}) and $|q_2^2|_{\max}\sim
\min \{{\bf k}_\perp^2/(1-y_2),\, W^2\}$.
At small transverse momentum of the pair $k_\perp$ this result
coincides with that obtained in Ref.~\cite{CS} (see Appendix~\ref{app:C}).
The contribution $d\sigma_{13}$ is obtained  from $d\sigma_{12}$
using replacements~(\ref{23}).

The two--photon and bremsstrahlung backgrounds can be found in
review \cite{BFKK}.

We have analyzed the charge asymmetry of muons in the same terms
as it was done for the pions. Table~\ref{tab:3}
\begin{table}[!htb]
  \begin{center}
  \caption{Muon pair charge asymmetry signals and background at
           different c.m. energies}
  \label{tab:3}
  \vspace{3mm}
  \begin{tabular}{ccccc}\hline
    $\sqrt{s}      $, GeV&  1 &    4  &   10 & 200   \\\hline
    $\Delta\sigma_K$, pb &-11 &  7.9  &   42 &  120  \\
    $\Delta\sigma_v$, pb &180 &  660  &  950 & 1700  \\
    $\Delta\sigma_B      $, pb &7100& 3.7 $\times10^4$
    &6.9 $\times 10^4$ & 3.7 $\times 10^5$
    \\\hline
  \end{tabular}
  \end{center}
\end{table}
contains
values of integrated signals and background at different c.m.
energies. From that table we observe:\\
(i) the muon transverse asymmetry $\Delta\sigma_v$ is considerably
larger than the muon longitudinal asymmetry $\Delta\sigma_K$;\\
(ii) the transverse asymmetry for muons is considerably
larger than that for pions (see Table~\ref{tab:1});\\
(iii) the longitudinal asymmetries for pions and muons are
of the same order of magnitude.

\begin{figure}[!htb]
  \centerline{\epsfig{file=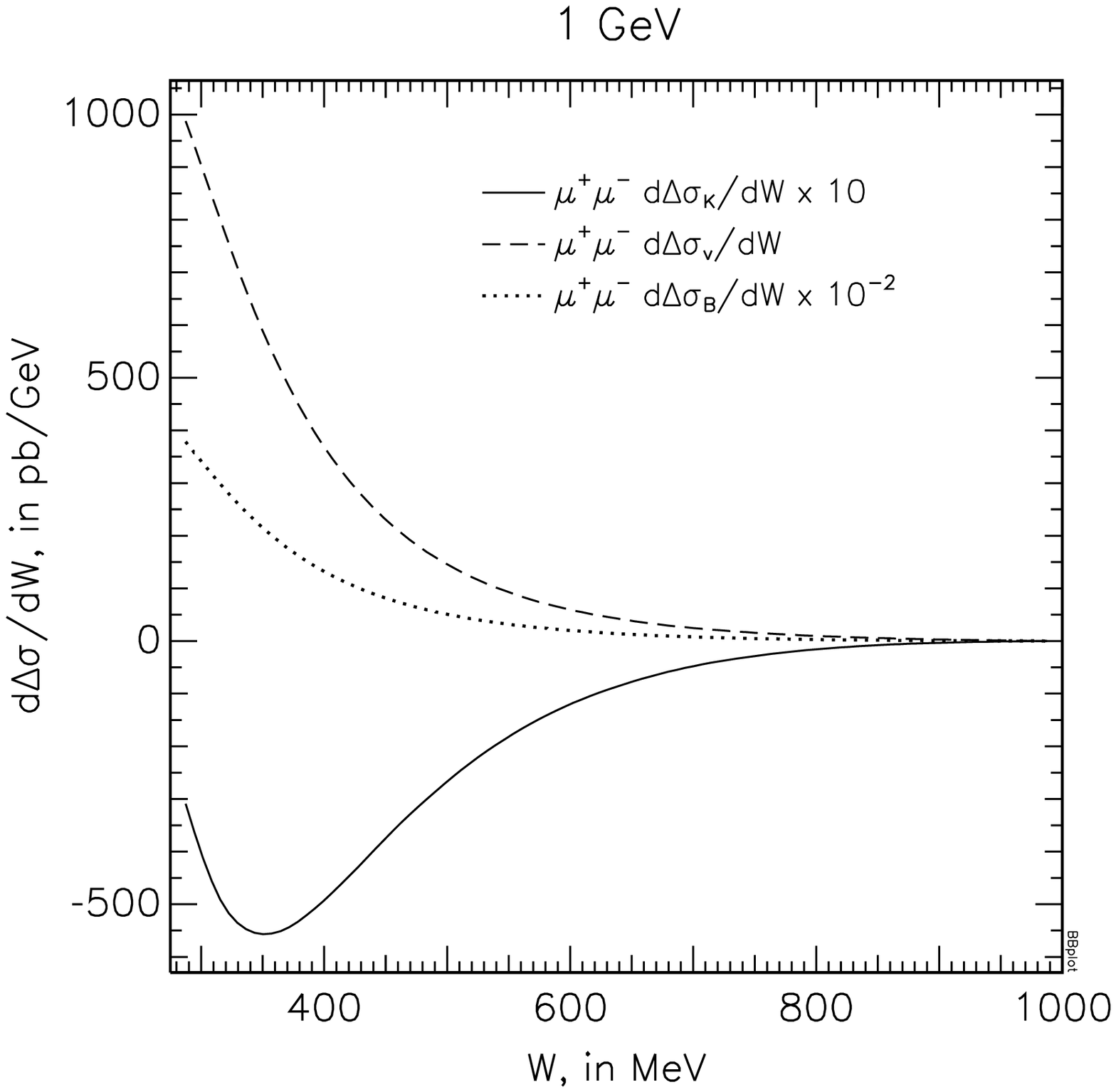,width=6cm}}
  \centerline{\epsfig{file=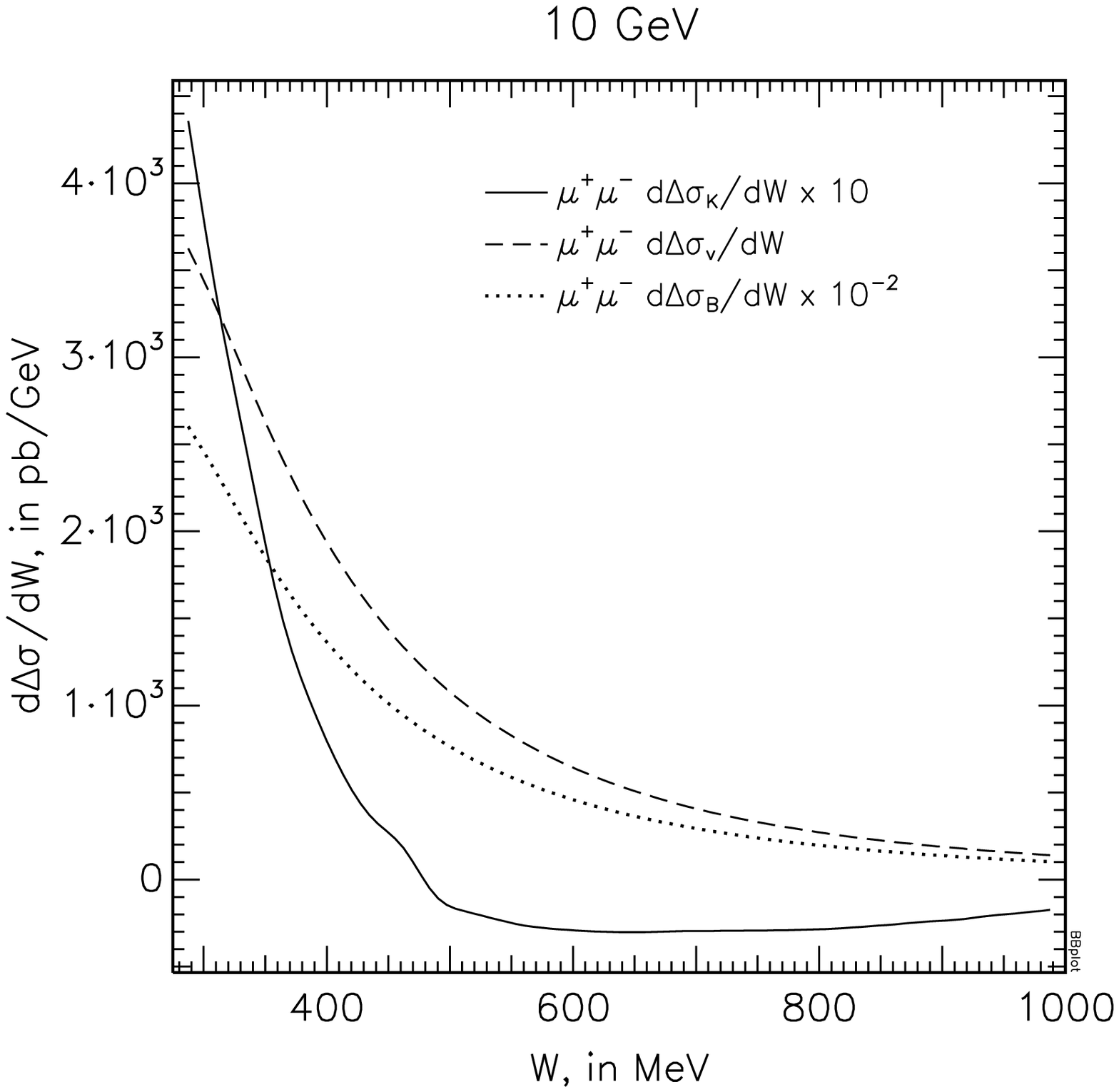,width=6cm}}
  \caption{
           Contributions $\Delta \sigma_K$ and $\Delta \sigma_v$
           (\protect\ref{sigw}) and background at
           $\protect\sqrt s = 1$ and 10 GeV vs. $W$
           for muon pair production.
          }
  \label{fig:9}
\end{figure}

\begin{figure}[!htb]
  \centerline{\epsfig{file=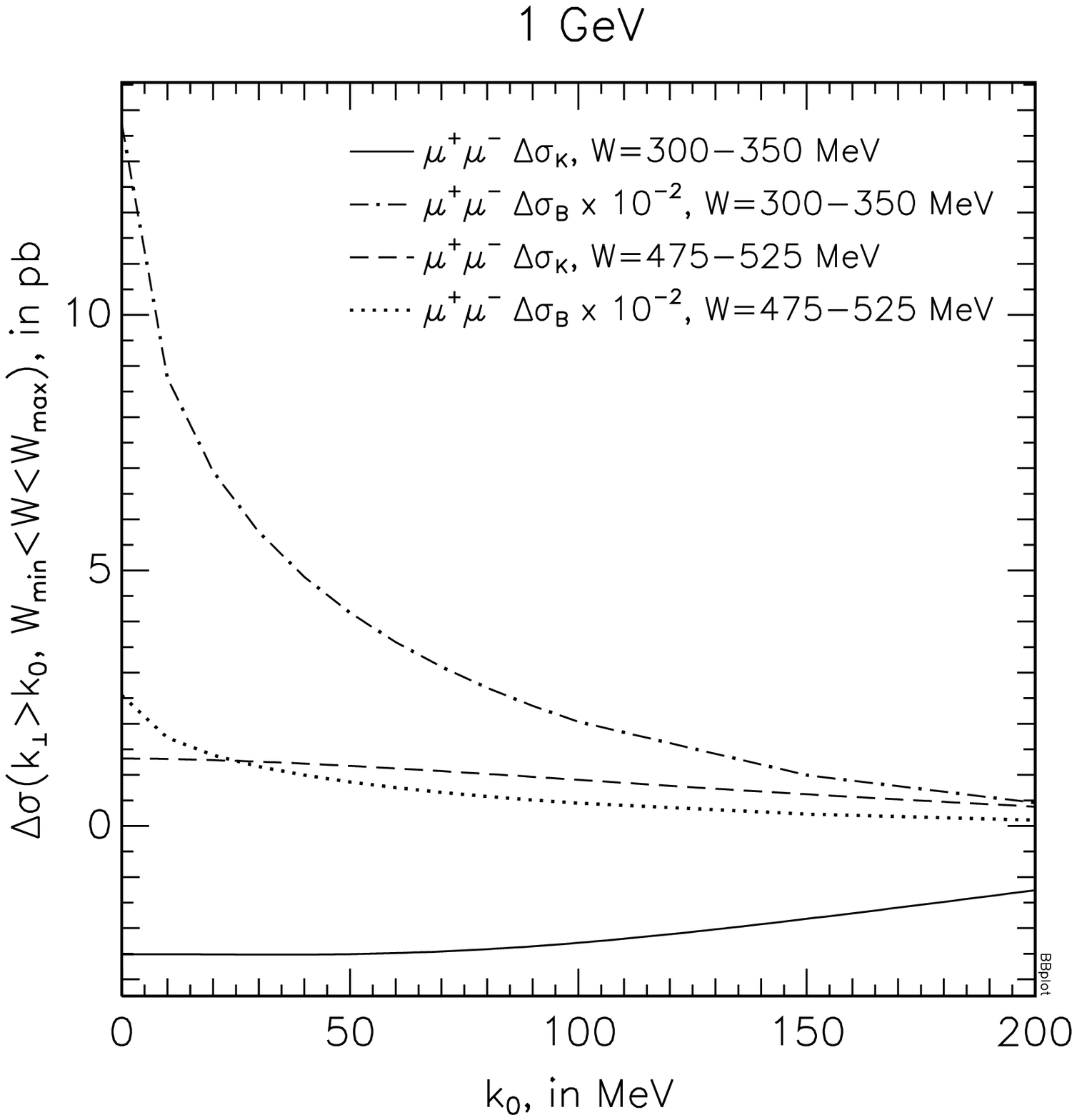,width=6cm}}
  \centerline{\epsfig{file=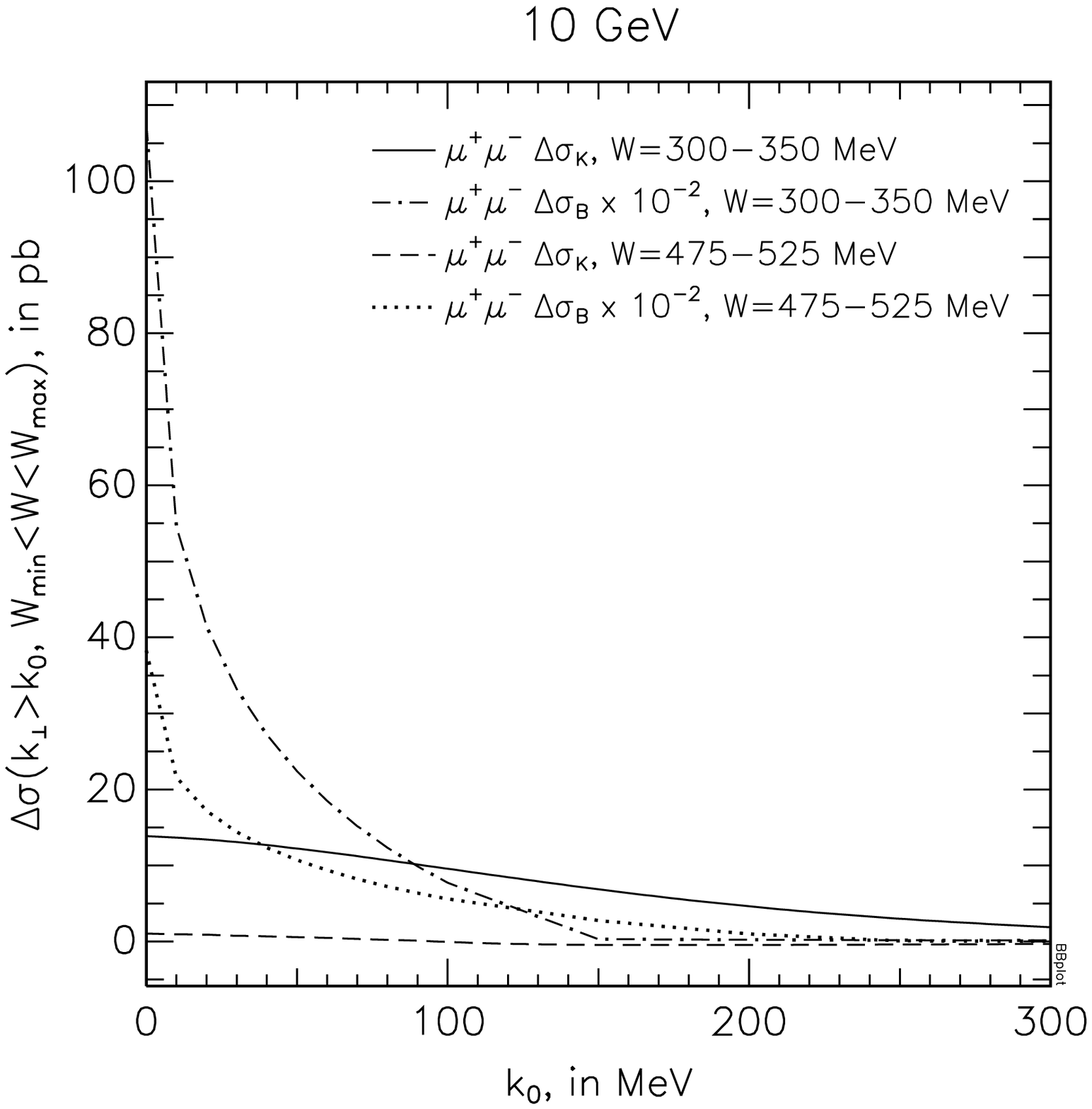,width=6cm}}
  \caption{
           The contribution $\Delta \sigma_K$ and background at
           $\protect\sqrt s  =1$ and 10 GeV for the intervals $W=
           300 \div 350$ MeV and $W=475 \div 525$ MeV and $k_\perp
           > k_0$ vs.  $k_0$ for muon pair production.
          }
  \label{fig:10}
\end{figure}

Similar relations between muon and pion asymmetries take
place in the all considered regions of parameters. Moreover,
in some regions the longitudinal asymmetry of muons disappear contrary
to that that of pions.
In Figs.~\ref{fig:9}---\ref{fig:10}
we present the distributions similar to those given in
Figs.~\ref{fig:4}---\ref{fig:5} for pions.  Having in mind the
muon production as a possible background for the pion production,
we consider just the same regions of $W$ as used for pions.
It is interesting to note that at $\sqrt{s} = 10$ GeV the
longitudinal asymmetry $d\Delta\sigma_K/dW$ (Fig.~\ref{fig:9})
changes its sign at $W\approx 500$ MeV. The position of this
crossover depends on  $\sqrt{s}$ (cf. the curve for $\sqrt{s}=1$
GeV). Such a sign change is absent for pions.

\section{Summary and conclusions}

We have calculated the charge asymmetry contribution $d\sigma_{\rm
interf}/(d^3 p_+d^3 p_-)$ for the process $e^+e^- \to e^+ e^-
\pi^+ \pi^-$. Our result summarized in Eqs.~(\ref{7}),(\ref{35}),(\ref{23}) is
expressed via the helicity amplitudes $M_{++}$, $M_{+-}$ and
$M_{0+}$ of the subprocess $\gamma^*\gamma\to \pi^+\pi^-$ and the pion
form factor $F_\pi$ in analytical form. It can be used for experiments
both without recording the
scattered electrons (no tag experiments) and with
recording one scattered electron (single tag experiments).

The experimentally observable effects can be studied,
in principle, in the
longitudinal asymmetry (in the variable $K_-$ of Eq.~(\ref{13})
representing the
difference in longitudinal momenta of positive and negative pions)
and in the transverse asymmetry (in the variable $v$ of Eq.~(\ref{13})
representing the difference between squared transverse momenta of
positive and negative pions). The
longitudinal asymmetry effects are saturated mainly by the
contribution of the $\ggam\to
\pi^+\pi^-$ helicity amplitude $M_{++}$. The main
contribution to the transverse asymmetry is given by amplitudes
$M_{++}$ and $M_{+-}$. The two--photon amplitude $M_{0+}$ (one
longitudinal (scalar) photon and one transverse photon) gives
typically about 10\% of effect (within the QED model).

The main part of our numerical analysis of the effect is performed in
the QED model giving a reasonable description of the two--photon
amplitude at $W\lsim 1$ GeV and the one--photon amplitude near the
threshold.

We have considered also the charge asymmetry for muon pairs which
can give an essential background to the
pion asymmetry effects. Our analysis demonstrates that the transverse
asymmetry of pions is much smaller than that of muons whereas the
longitudinal asymmetry of pions is generally close to that of
muons (in some regions of parameters the latter even disappears).
Therefore, in our detailed numerical studies we have concentrated our
efforts to the study of the longitudinal asymmetry.

Our QED numerical analysis presents a good basis to estimate
the potentiality for studying pion--pion scattering phase shifts
near the threshold at the colliders DA$\Phi$NE, VEPP2000, etc. The
study of resonances (for example, different $f_0$'s and $f_2$) at
the colliders PEP-II and KEK-B requires more detailed estimates
which are in progress.

The values of the signal to background ratio $S/B$ (\ref{SB})
obtained in the QED model are typically around 1\%. However, with the
expected high luminosity of B-- and $\phi$--factories the
statistical significance of the effect $SS$ (\ref{SS}) is high enough
(typically about 10) for the considered ideal case when all
dipions are assumed to be recorded.
The strongly different dependence of
signal and background on the components of the total dipion
momentum results in a large
 improvement of both $S/B$ and $SS$ (by
about a factor 10 for $W\sim 0.5\div 1$ GeV at the PEP-II) at
suitable cuts in the total transverse momentum of pion pair
$k_\perp$ and its total rapidity (see Table~\ref{tab:2}). With these
estimates the perspectives of experiments look well.

The charge asymmetry in the resonance region will be considered
elsewhere. Preliminary estimates
indicate that the effect will be observable. Moreover,
the quick change of the phase near the resonance results in a
strong dependence of the asymmetry on $W$
in this region what could help to
distinguish resonance models.

Finally note  that the presented equations for the muon charge
asymmetry (\ref{38b}) can be used to estimate the
 production of
$c$ quarks in the process $e^+e^-\to e^+e^-c\bar{c}$ with
production of open charm. The deviation from the QED result will
be caused by violation of quark--hadron duality due to strong
interactions. These deviations are expected to be large near the
threshold where the $c\bar{c}$ resonance production is essential
in both $p$-- and
$s$--waves. Simple QED calculations with an additional
charge factor 8/9
give  $\Delta\sigma_v\sim$ 0.27 pb at
$\sqrt{s}=10$ GeV and about 4.1 pb at $\sqrt{s}=200$ GeV
(here we put $c$--quark mass of 1.75 GeV to take into account a
threshold for the production of the open charm).

\begin{acknowledgement}
We are very grateful to A.~Bondar, V.~Fadin, I.~Ivanov, E.~Kuraev,
V.~Savinov, V.~Serebryakov, W.~Speth and V.~Zhilich for useful
discussions.  This work was partially supported by grants RFBR
00-02-17592, 00-15-96691, ``Universities of Russia"  015.0201.16,
and Sankt-Petersburg Center of Higher education.
V.G.S. acknowledges a support of S\"achsisches Staats\-ministerium
f\"ur Kunst und Wissenschaft, grant 4-7531.50-04-0361-00/25
and kind hospitality of NTZ of Leipzig university.
\end{acknowledgement}

\begin{appendix}

\section{Kinematics of the process $\gamma^*\gamma^* \to \pi^+ \pi^-$}
\label{app:A}
\appeqn

Due to 4--momentum conservation $q_1 +q_2 = p_+ + p_-$, the
amplitude $M^{\mu \nu}$ of the process $\gamma^*\gamma^* \to
\pi^+ \pi^-$ depends only on three independent momenta $q_1,\;
q_2, \; \Delta=p_+ -p_-$. We use the Mandelstam variables
$$
  W^2 =(q_1+q_2)^2,\;\; t=(q_1-p_+)^2,\;\; u= (q_1-p_+)^2
$$
with the relations
$$
  W^2+ t+u = 2\mu^2 +q_1^2+q_2^2\,,\;\;
  t-u = -2q_1 \Delta = 2q_2 \Delta\,.
$$

The polarization properties of the virtual photons are described
by two polarization 4--vectors $e^{(a)}_1$ and $e^{(b)}_2$ with
helicities $a,b =\pm 1,\; 0$.  For scalar (longitudinal) photons
($a=b=0$) we use the following polarization vectors
\bea
  e^{(0)}_{1\mu} &=& \sqrt{\fr{-q_1^2 }{X}} \left( q_{2\mu}- \fr{q_1q_2
  }{ q_1^2}\, q_{1\mu} \right)\,,
  \nonumber \\
  e^{(0)}_{2\nu} &=& \sqrt{\fr{-q_2^2}{X}} \left( q_{1\nu}- \fr{q_1q_2
  }{ q_2^2}\, q_{2\nu} \right)\,,
  \label{A.1}
  \\
   X &=& (q_1 q_2)^2 -q_1^2 q_2^2\,.
  \nonumber
\eea
The transverse photons ($a, b =\pm 1$) can be described by two
independent polarization vectors only.  This choice can be
realized by taking those vectors in the $\gamma^* \gamma^*$
center--of--mass system in the form
\be
  e^{(\pm)}_{1} = e^{(\mp)}_{2}=
  \mp \, \fr{1}{\sqrt{2}}\, (0,\;1,\;\pm {\rm i},\; 0).
  \label{A.2}
\ee

To construct the corresponding covariant expression,
we
introduce a metric tensor of a subspace which is orthogonal to
the 4--vectors $q_1$ and $q_2$
\begin{eqnarray*}
  R_{\mu\nu} &=& g_{\mu\nu}-
  \\
  &-&\! \fr{1}{X}
  \left[ (q_1q_2)\left( q_{1\mu} q_{2\nu}+ q_{1\nu}
  q_{2\mu}\right) - q_1^2 q_{2\mu} q_{2\nu} -
  q_2^2 q_{1\mu} q_{1\nu} \right]
\end{eqnarray*}
and define two 4--vectors in that subspace (both of them are
antisymmetric under  $p_+ \leftrightarrow p_-$ exchange)
\bea
  r^\mu = \fr{1}{2} \, R^{\mu \nu}\, \Delta_\nu\,, \quad
  s^\mu &=& \varepsilon^{\mu \nu \alpha \beta} \Delta_\nu\, q_{1\alpha}\,
  q_{2\beta}\,,
  \nonumber \\
  r^\mu s_\mu&=&0\,.
  \label{A.3}
\eea
In the $\gamma^*\gamma^*$ c.m.s. the tensor $R_{\mu\nu}$ has only two nonzero
components $R_{xx}=R_{yy}=-1$. Both 4--vectors $r^\mu$ and $s^\mu$
have only nonzero components in $x$ and $y$ directions and are
perpendicular to each other\footnote{In the $\gamma^* \gamma^*$
c.m.s. the nonzero components of the 4-vector $r$ coincides with
the transverse components of the vector ${\bf p}_{+\perp}$ or with
the transverse components of the vector $(-{\bf p}_{-\perp})$.}.
Therefore, we can choose the $x$-- and $y$--axes along these 4--vectors:
\be
  e^{(x)}_\mu =\fr{r_\mu }{ \sqrt{-r^2}}\,,\qquad
  e^{(y)}_\mu = \fr{s_\mu }{ \sqrt{-s^2}}\,.
  \label{A.4}
\ee
In the $\gamma^* \gamma^*$ c.m.s. $e^{(x)} = (0,\; 1,\;0,\;0)$,
$e^{(y)} = (0,\; 0,\;1,\;0)$. As a result, the covariant expression for the
vectors (\ref{A.2}) takes the form
\be
  e^{(\pm)}_{1\mu} =
  e^{(\mp)}_{2\mu}= \mp\,\fr{1}{ \sqrt{2}}\, \left(e^{(x)}_\mu \pm
  {\rm i} e^{(y)}_\mu\right)\,.
  \label{A.5}
\ee

We define the helicity amplitudes for the discussed process as
$$
  M_{ab}= e_{1\mu}^{(a)}\, e_{2\nu}^{(b)}\, M^{\mu \nu}\,,
$$
where the inverse transformation is given in Eq.~(\ref{16}).
Taking into account parity conservation, we obtain
\bea
  M_{++}&=&M_{--}\,,\qquad M_{+-}=M_{-+}\,,
  \nonumber \\
  M_{0+}&=&-M_{0-}\,,\qquad M_{+0}=-M_{-0}\,.
  \label{A.6}
\eea
Since under $\pi^+\leftrightarrow \pi^-$ exchange
the $C$--even
amplitude $M^{\mu \nu}$ is symmetric whereas the 4--vectors
$e_{1,2}^{(\pm)}$ are antisymmetric [cf. Eq.~(\ref{Cact})], 
the amplitudes $M_{++}$,
$M_{+-}$ and $M_{00}$ are symmetric and  the amplitudes $M_{0+}$
and $M_{+0}$ are antisymmetric under that exchange
[see Eq.~(\ref{C2prop})].

Under photon exchange ($q_1 \leftrightarrow q_2$) the
polarization vectors have to replaced by $e^{(0)}_1 \to
e^{(0)}_2$, $e^{(x)} \to e^{(x)}$, $e^{(y)} \to \,-\, e^{(y)}$,
$e^{(\pm)}_1 \to\,-\, e^{(\pm)}_2$ what can be short-hand
written as
$$
  e^{(a)}_1 \to \, (-1)^a\, e^{(a)}_2\,.
$$
Taking into account $M^{\mu \nu} (q_1,\, q_2, \Delta)=M^{\nu
\mu} (q_2,\, q_1, \Delta)$, we obtain
\be
  M_{ab}(q_1,\,q_2,\, \Delta)\, \to \,(-1)^{a+b}\,
  M_{ba}(q_1,\,q_2,\, \Delta)\,.
  \label{A.7}
\ee

It is useful to present  these amplitudes for the pure QED case
(point--like pions)
\bea
  M_{++}^{QED}&=&8\pi\alpha - M_{+-}^{QED}\,,
  \nonumber \\
  M_{+-}^{QED}&=&-8\pi\alpha
  \fr{r^2(W^2-q_1^2-q_2^2)}{(t-\mu^2) (u-\mu^2)}\,,
  \label{A.8}
  \\
  M_{+0}^{QED}&=&-2\pi\alpha\sqrt{\fr{2q_2^2 r^2}{X}} \, \fr{(t-u)
  (W^2+q_2^2-q_1^2)} {(t-\mu^2)(u-\mu^2)}\,,
  \nonumber \\
  M_{0+}^{QED}&=& -2\pi\alpha\sqrt{\fr{2q_1^2 r^2}{X}} \,
  \fr{(t-u)(W^2+q_1^2-q_2^2)} {(t-\mu^2)(u-\mu^2)}\,.
  \nonumber
\eea
with
\be
  r^2 = -\fr{W^2}{ 4X}\, \left[ (t-\mu^2)(u-\mu^2) -q_1^2 q_2^2
  \right] +\mu^2 < 0.
  \label{A.9}
\ee
In the $\gamma^* \gamma^*$ cms the quantity $(-r^2)$ is the
squared transverse momentum of $\pi^+$ or $\pi^-$.

\section{Calculation of traces}
\label{app:B}
\appeqn

\subsection{Sudakov variables}

When calculating the trace (\ref{22}), we neglect the electron
mass $m_e$ and decompose any 4--vector $A$ into components in
the plane of the 4--vectors $p_1$ and $p_2$ and in the plane
orthogonal to them
\bea
  A&=&x_Ap_1+y_Ap_2+A_\perp\,,
  \label{B.1}
  \\
  x_A&=&\fr{2p_2A}{s}\,,\quad
  y_A=\fr{2p_1A}{s}\,,\quad
  A^2 = s\,x_A y_A + A_\perp^2\,.
  \nonumber
\eea
The parameters $x_A$ and $y_A$ are so called Sudakov variables;
in the collider system described in Sect.~\ref{sec:bas} the 4--vector
$A_\perp$ has $x$ and $y$ components only
\be
  A_\perp = (0,\; A_x,\;A_y,\;0)\,,\;\;
  A_\perp^2 =-{\bf A}_\perp^2\,.
  \label{B.2)}
\ee
In particular,
\bea
  p_\pm&=&x_\pm p_1+y_\pm p_2+p_{\pm \perp}\,,
  \nonumber \\
  \Delta &=&x_\Delta p_1 + y_\Delta p_2 + \Delta_\perp\,,
  \label{B.3}
  \\
  q_i&=&x_ip_1+y_ip_2+q_{i \perp}
  \nonumber
\eea
with $x_{\pm}$, $y_{\pm}$, $x=x_+ +x_-$ and $y=y_+ +y_-$
mentioned in Eq.~(\ref{12}).

In the used logarithmic approximation [see Eq.~(\ref{26})] the
decomposition of $q_2$ and  $r$ defined in Eq.~(\ref{A.3}) is
simplified
\bea
  q_2 &=& y_2 p_2, \quad
  r=y_r p_2 + r_\perp,
  \label{B.4}
  \\
  r^2 &=& -{\bf r}_\perp^2 \,, \quad
  y_r = \frac{2}{x}\, {\bf r}_\perp {\bf k}_\perp
  \nonumber
\eea
with
$y_2$ and ${\bf r}_\perp$ given in Eqs.~(\ref{28}) and
(\ref{31}).
Besides,
\begin{eqnarray*}
  t-\mu^2 &=& -2q_2 p_- = -sy_2 x_-\,,
  \\
  u-\mu^2 &=& -2q_2 p_+ = -sy_2 x_+\,,
  \\
  t-u &=& 2q_2 \Delta = sy_2 \xi x\,.
\end{eqnarray*}
Below we use the notations
\be
  e^{(a)} \equiv e_1^{(a)}=x^{(a)}p_1+y^{(a)}p_2+e^{(a)}_\perp
  \label{B.5}
\ee
with
\bea
  x^{(\pm)}&=&0\,,\quad y^{(\pm)}= \fr{2 {\bf e}^{(\pm)}_\perp{\bf
  k}_\perp }{sx}\,,
  \nonumber \\
  y^{(0)}&=&\sqrt{-q_1^2}\fr{(2-x)}{xs}\,,\quad
  {\bf e}_\perp^{(0)}=\fr{{\bf k}_\perp}{ \sqrt{-q_1^2}}\,.
  \label{B.6}
\eea
The normalization condition for the 4--vectors $e^{(a)}_{\mu}$
results in
$$
  {\bf e}^{(a)*}_\perp {\bf e}^{(b)}_\perp=\, \delta_{ab}\,,\quad
  a\,, b = \pm 1\,.
$$
The following expressions will be useful ($i,j=x,y$):
\be
  \left({\bf e}^{(+)*}_\perp \right)_i
  \left({\bf e}^{(+)}_\perp \right)_j + \left({\bf e}^{(-)*}_\perp
  \right)_i \left({\bf e}^{(-)}_\perp \right)_j = \delta_{ij}\,,
  \label{B.7}
\ee
\bea
  &&\left({\bf e}^{(+)*}_\perp \right)_i
  \left({\bf e}^{(-)}_\perp \right)_j + \left({\bf e}^{(-)*}_\perp
  \right)_i \left({\bf e}^{(+)}_\perp \right)_j
  = \delta_{ij} -
  2  e^{(x)}_i\, e^{(x)}_j \,,
  \nonumber \\
  && \label{B.8}
\eea
\be
  \left({\bf e}^{(+)}_\perp \right)_i -
  \left({\bf e}^{(-)}_\perp \right)_i
  =-\sqrt{2}  e^{(x)}_i\,,
  \label{B.9}
\ee
where the vector ${\bf e}^{(x)}_\perp$ has the form [in
accordance with definition (\ref{A.4})]
\be
  {\bf e}^{(x)}_\perp = \fr{ {\bf r}_\perp }{|{\bf r}_\perp|}\,.
  \label{B.10}
\ee

\subsection{Calculation of $C_1^{ab}$}

Now we describe the calculation of trace (\ref{22})   in the
used approximation (\ref{26}).  First, we rewrite the trace
$C_1^{ab}$ with $a=0,\;\pm 1$ and $b= \pm 1$ in the form of
three terms
\bea
  C_1^{ab}&=&\fr{(-1)^{a+1}}{2} \times
  \nonumber \\
  &{\mathrm{Tr}}&\left[\hat{p}'_1
  \hat{e}^{(a)*}\hat{p}_1 \left(\hat{e}^{(b)}
  \fr{\hat{p}_1+\hat{q}_2}{sy_2}\hat{\Delta}-
  \hat{\Delta}\fr{\hat{p}_1^\prime-\hat{q}_2}{sy_2(1-x)}
  \hat{e}^{(b)}\right)\right]
  \nonumber \\
  &=&\fr{(-1)^{1+a}}{2 sy_2(1-x)}\left[N^{ab}_1+y_2\left(N^{ab}_2
  +N^{ab}_3\right)\right]\,,
  \label{B.11}
  \\
  N^{ab}_1&=&{\mathrm{Tr}}\left[\hat{p}_1^{\prime}
  \hat{e}^{(a)*}\hat{p}_1
  \left((1-x)\hat{e}^{(b)}\hat{p}_1\hat{\Delta}-
  \hat{\Delta}\hat{p}_1^\prime
  \hat{e}^{(b)}\right)\right]\,,
  \nonumber \\
  N^{ab}_2&=&(1-x){\mathrm{Tr}}\left[\hat{p}_1^{\prime}
  \hat{e}^{(a)*}\hat{p}_1
  \hat{e}^{(b)}\hat{p}_2\hat{\Delta}\right]\,,
  \nonumber \\
  N^{ab}_3&=&{\mathrm{Tr}}\left[\hat{p}_1^{\prime}\hat{e}^{(a)*}
  \hat{p}_1 \hat{\Delta}\hat{p}_2\hat{e}^{(b)}\right]\,.
  \nonumber
\eea
The polarization 4--vector $e^{(a)}$ appears in $C_1^{ab}$ in the
combination  $\hat{e}^{(a)*} \hat{p}_1$ only, therefore, the
$p_1$ component of ${e}^{(a)}$ does not contribute ($\hat p_1
\hat p_1 \approx 0$).  Since $x^{(\pm)}=0$, such a component is
also absent in  $e^{(b)}$.  As a result, we can use all
polarization 4--vectors in the form
$$
  e^{(a)} =y^{(a)}p_2+e^{(a)}_\perp\,.
$$

Next, for $N_1$ we use the relations
\begin{eqnarray*}
  \hat{p}_1\hat{e}^{(b)}\hat {p}_1&=&sy^{(b)}\,\hat{p}_1\,,
  \\
  \hat{p}'_1\hat{e}^{(b)}\hat {p}'_1&=&2(e^{(b)}p'_1)\hat{p}'_1
  = [(1-x) sy^{(b)} +2({\bf k_\bot}{\bf
  e}^{(b)}_\bot)]\hat{p}'_1
\end{eqnarray*}
which results in
$$
  N_1^{ab}=-2({\bf k_\bot}{\bf e}_\bot^{(b)})\,
  {\mathrm{Tr}}\,\left[\hat{p}_1^{\prime}\hat{e}^{(a)*}\hat{p}_1
  \hat{\Delta}\right]\,.
$$
For the trace we obtain
\bea
  {\mathrm{Tr}}\,
  &&\left[\hat{p}_1^{\prime}\hat{e}^{(a)*}\hat{p}_1 \hat{\Delta}\right]=
  2 s
  \left[s(1-x)y^{(a)*}y_\Delta + \right.
  \nonumber \\
   && \left.
  +\fr{-q_1^2}{s}({\bf\Delta}_\bot{\bf
  e}_\bot^{(a)*}) +y_\Delta({\bf k}_\bot{\bf e}_\bot^{(a)*}) +
  y^{(a)*} ({\bf \Delta}_\bot{\bf k}_\bot)\right]\,.
  \nonumber
\eea
The final result can be expressed in the form
\bea
  N_1^{ab}=-4s
  \left[y_\Delta D_3 +\fr{-q_1^2}{s}D_2 +\left((1-x)y_\Delta
  +\fr{v}{s}\right)D_4\right]
  \nonumber
\eea
where we have used the basic structures
\bea
  D_0&=&({\bf e}_\bot^{(a)*}{\bf e}_\bot^{(b)})\,,\quad
  D_1=({\bf e}_\bot^{(a)*}{\bf k}_\bot)({\bf e}_\bot^{(b)}{\bf
  \Delta}_\bot)\,,
  \nonumber \\
  D_2&=&({\bf e}_\bot^{(a)*}{\bf \Delta}_\bot)
  ({\bf e}_\bot^{(b)}{\bf k}_\bot)\,, \quad
  D_3=({\bf e}_\bot^{(a)*}{\bf k}_\bot)({\bf e}_\bot^{(b)}{\bf k}_\bot)\,,
  \nonumber \\
  D_4&=&sy^{(a)*}({\bf e}_\bot^{(b)}{\bf k}_\bot)\,, \quad
  D_5=sy^{(a)*}({\bf e}_\bot^{(b)}{\bf \Delta}_\bot)\,.
  \nonumber
\eea

Similarly we obtain
\bea
  N_2^{ab}&=&2s(1-x)\left[(v-q_1^2x_\Delta)D_0+D_1-\right.
  \nonumber \\
  && \left. -D_2+x_\Delta
  D_4+(1-x)D_5\right]\,,
  \nonumber \\
  N_3^{ab} &=&2s\left[-vD_0+D_1+D_2+(1-x)D_5\right]
  \nonumber
\eea
and the final expression for $C_1^{ab}$
\bea
  C_1^{ab}&=& \fr{(-1)^{1+a}}{y_2(1-x)}\left[-xy_2(v-\xi{\bf
  k}_\bot^2)D_0 + (2-x)y_2D_1 +
  \right.
  \nonumber \\
  &+& \left. \left(\fr{2q_1^2}{s}+xy_2\right)D_2
  -2y_\Delta D_3- \right.
  \label{B.12}
  \\
  &-&\left. \left(\fr{2v}{s}+2(1-x) y_\Delta-y_2x\xi (1-x)\right)D_4+
  \right.
  \nonumber \\
  &+&\left.y_2(1-x)(2-x)D_5\right]\,.
  \nonumber
\eea
with
$$
  sx y_\Delta= 2x(p_1 \Delta) = 2v-\xi (W^2+{\bf k}_\perp^2)\,.
$$

\subsection{Calculation of $T=\sum\limits_{ab} M_{ab}C_1^{ab}$}

Using Eq.~(\ref{B.6}) we find for $T$ the expression
\bea
  T&=&(C_1^{++}+C_1^{--})M_{++}+(C_1^{+-}+C_1^{-+})M_{+-}+
  \nonumber \\
  &&+(C_1^{0+}-C_1^{0-})M_{0+}
  \label{B.13}
  \\
  &=&\fr{2}{y_2(1-x)} [G^0 M_{++} + G^2M_{+-}
  + G^1M_{0+}]\,.
  \nonumber
\eea
Taking into account Eq.~(\ref{B.7}) the basic structures related to the
amplitude $M_{++}$ are
\bea
  D_0&=& 2\,,\quad
  D_1= D_2= v\,,\quad
  D_3={\bf k}_\bot^2\,,
  \nonumber \\
  D_4 &=& \fr{2}{x} D_3 \,, \quad
  D_5 = \fr{2}{x} D_1
  \nonumber
\eea
which leads to the coefficient $G_0$
\bea
  G^0&=&\fr{v}{x} \left[ 2(1-x)
  y_2 - \fr{(2-x)}{ (1-x)} \fr{{\bf k}_\perp^2}{s}\right] +
  \nonumber \\
  &+&
  {\bf k}_\perp^2  \left[ y_2 \xi - \fr{(2-x) y_\Delta}{x}
  \right]\,.
  \label{B.14}
\eea
Analogously we
get for the basic structures related to $M_{+-}$
[taking into account Eq.~(\ref{B.8})]
\begin{eqnarray*}
  D_0&=& 0\,,\quad
  D_1= D_2= D_1^{(2)}=v- \fr{(v-\xi{\bf k}_\perp^2)
  ({\bf \Delta}_\perp^2-\xi v)}{2{\bf r}_\perp^2}\,,
  \\
  D_3&=& D_3^{(2)}={\bf k}_\bot^2-\fr{(v-\xi{\bf k}_\bot^2)^2}
  {2{\bf r}_\perp^2}
  \,,
  \\
  D_4&=&\fr{2}{x}D_3^{(2)} \,,\quad D_5 =\fr{2}{x}D_1^{(2)}
\end{eqnarray*}
which gives the coefficient $G_2$:
\bea
  G^2&=& \left[ \fr{2 y_2}{x} \left(1-x+\fr{x^2}{2}\right) -
  \fr{{\bf k}_\perp^2}{(1-x) s}\right]D_1^{(2)} +
  \nonumber \\
  &+&
  \left[ (1-x)y_2\xi - \fr{(2-x) y_\Delta}{x}  -
  \fr{2v}{ xs} \right]D_3^{(2)}.
  \label{B.15}
\eea
Finally the coefficient $G^1$ is obtained from [using (\ref{B.9})]
\begin{eqnarray*}
  D_0&=& D_0^{(1)}= -
  \fr { v-\xi {\bf k}_\bot^2 }   {\sqrt{-2q_1^2
  {\bf r}_\perp^2}} \,,
  \\
  D_1&=& D_1^{(1)}= -\fr{({\bf \Delta}_\perp^2-\xi v) {\bf
  k}_\bot^2}{\sqrt{-2q_1^2{\bf r}_\perp^2}}\,,
  \\
  D_2&=& vD_0^{(1)}\,,\quad
  D_3= {\bf k}_\bot^2D_0^{(1)}\,
  \\
  D_4&=&\fr{2-x}{x(1-x)} {\bf k}_\bot^2 D_0^{(1)}\, \quad
  D_5=\fr{2-x}{x(1-x)}D_1^{(1)}
\end{eqnarray*}
as follows
\bea
  G^1 &=& -{\bf k}_\perp^2 \left[ y_2 \xi - \fr{2 y_\Delta}{x} -
  \fr{2v}{x(1-x)s} \right]D_0^{(1)} -
  \nonumber \\
  &-&\fr{2-x}{x} y_2 D_1^{(1)} \,.
  \label{B.16}
\eea

The coefficients $G^n$  depend on the vectors ${\bf
\Delta}_\perp, \; {\bf k}_\perp$ and ${\bf r}_\perp$. Using the
relation ${\bf \Delta}_\perp = 2{\bf r}_\perp +\xi {\bf
k}_\perp$ [see Eq.~(\ref{31})] and introducing the angle $\phi$
between the vectors ${\bf r}_\perp$ and ${\bf k}_\perp$ we get
after some algebra the compact expression for $T$ in the form of
Eqs.~(\ref{7},\ref{35}).

\section{Some useful notes}
\label{app:C}
\appeqn

\subsection{Substitution rule for the contribution $d\sigma_{13}$}

Let us briefly describe  the necessary changes for
$d\sigma_{13}$ in the case of electron--positron or
electron--electron collisions
$$
  e^- (p_1) + e^{\pm} (p_2) \to e^-
  (p_1^\prime) + e^{\pm} (p_2^\prime) +\pi^+ \pi^-\,.
$$
The contribution $d\sigma_{13}$ has the form
\begin{eqnarray*}
  d\sigma_{13} &=& 2\,{\rm Re}({\cal M}_3^* {\cal
  M}_1) \, \fr{d\Gamma }{ 2s}
  \\
  &=& -2\, \fr{(4\pi \alpha)^3 }{ q_1^2
  q_2^2 k^2}\, \sum\limits_{abc=\pm 1, \,0}\, {\rm Re} \,\left(
  F_\pi^* \, M_{ab}\, \varrho_1^{ac}\, C_2^{bc}\, \right) \,
  \fr{d\Gamma}{ 2s}
\end{eqnarray*}
where $\varrho_1^{ac}$ and $C_2^{bc}$ are similar to
$\varrho_2^{bc}$ and $C_1^{ac}$ in Eqs.~(\ref{21}), (\ref{22}).
$d\sigma_{13}$ can be obtained from $d\sigma_{12}$ under the
exchange
$$
  p_1 \leftrightarrow  p_2\,, \; p_1^\prime
  \leftrightarrow  p_2^\prime\,, \; q_1 \leftrightarrow q_2\,.
$$
It is not difficult to check that in that case
\be
  \varrho_2^{bc} \to (-1)^{b+c}\, \varrho_1^{bc}\,,\;\;
  C_1^{ac} \to\,\mp\, (-1)^{a+c}\, C_2^{ac}\,.
  \label{C.1}
\ee
The sign $\mp$ in the last equation is in agreement with the
transition from three $e^-$ vertices in $C_1^{ac}$ to three
$e^{\pm}$ vertices in $C_2^{ac}$.

As a result, taking into account Eq.~(\ref{A.7}), we have
\be
  d\sigma_{13}\, = \, \mp \,
  d\sigma_{12} (p_1 \leftrightarrow p_2\,,\;
  p_1^\prime \leftrightarrow p_2^\prime\,,\;q_1 \leftrightarrow
  q_2\,)
  \label{C.2}
\ee
where the sign ``minus'' corresponds to electron--positron
collisions considered here and  ``plus''  to
electron--electron collisions.

\subsection{Absence of amplitude $M_{+-}$ in $d\sigma_{12}$
averaging over angle $\phi$}

The amplitude $M_{+-}$ enters the result (\ref{29})
with coefficient $C_1^{+-}+ C_1^{-+}$ [see Eq.~(\ref{B.13})]. Let us
consider $C_1^{-+}$  given by Eq.~(\ref{B.11}). Since
$$
  e^{(-)*} =- e^{(+)} \equiv - e
$$
and
$$
  eq_1=eq_2=0\,,\;\; ep_1 =ep_1^\prime\,, \;\;
  \hat{e} \hat{e}=0\,,
$$
we have
$$
  \hat{e} \hat{p_1} \hat{e} =
  \hat{e} \hat{p_1^\prime} \hat{e}  = 2(ep_1) \hat{e} \,.
$$
Therefore, the $C_1^{-+}$ coefficient can be simplified to a
trace of four Dirac matrices easily calculable
\begin{eqnarray*}
  C_1^{-+}&=&- (ep_1)\, {\mathrm{Tr}}
  \left\{ \hat{e}\, \fr{\hat{p}_1+\hat{q}_2}{sy_2} \,\hat{\Delta}
  \hat{p}'_1  -
  \hat{e} \hat{p_1} \hat{\Delta}\, \fr{\hat{p}_1-\hat{k}}{sy_2(1-x)}
  \right\}
  \\
  &=& \fr{(ep_1)^2\, f_1 + (ep_1)(e\Delta) f_2 }{ sy_2 (1-x)}
\end{eqnarray*}
with
\begin{eqnarray*}
  f_1 &=& 8x p_1 \Delta - 4(1-x)\, q_2 \Delta\,,
  \\
  f_2 &=& 4x p_1k + 4(1-x) \,q_2(p_1+p_1^\prime)\,.
\end{eqnarray*}
It is easy to check that only two scalar products depend on the
azimuthal angle $\phi$:
$$
  2ep_1 = sy^{(+)} = \fr{2}{ x}\, {\bf e}^{(+)} {\bf k}_\perp =
  -\fr{\sqrt{2}}{ x}\, |{\bf k}_\perp| {\mathrm e}^{{\mathrm i}
  \phi}\,,
$$
$$
 2 x p_1 \Delta= 4 |{\bf r}_\perp|\, | {\bf k}_\perp| \cos \phi -
 \xi \left( W^2- {\bf k}_\perp^2 \right) \,.
$$
As a result, the structure of $C_1^{-+}$ is
$$
  C_1^{-+}=  (a+b
  \cos{\phi})\, {\mathrm e}^{2{\mathrm i}\phi} + c \, {\mathrm
  e}^{{\mathrm i}\phi}
$$
which leads to
\begin{eqnarray*}
  C_1^{+-}+ C_1^{-+}&=& 2 {\rm Re}\, C_1^{-+}
  \\
  &=&
  (b+2c) \cos{\phi} +2a \cos{2\phi} +b \cos{3\phi}\,.
\end{eqnarray*}
Therefore, this coefficient disappears after averaging over  $\phi$
\be
  \langle \,C_1^{+-}+ C_1^{-+}\, \rangle_\phi =0\,.
  \label{C.3}
\ee

\subsection{Low $k_\bot$ limit for $d\sigma_{12}$}

At small transverse momentum of produced the pion pair $k_\bot$ our
result (\ref{29}),(\ref{7}) is simplified to
\bea
   \varepsilon_+ \varepsilon_-
  \fr{d\sigma_{12}}{d^3p_+\,d^3p_-}
  &=& - \fr{\alpha^3}{4\pi^4}\, \fr{x}{W^6 d}\,
  \fr{{\bf k}_\perp {\bf \Delta}_\perp}{{\bf k}_\perp^2} \times
  \nonumber  \\
  &&\left( 1-y_2 +\fr{1}{2} y^2_2 \right) \,L_2\,\times
  \label{C.4}  \\
  \Biggl[ (1-x)\,{\rm Re}\left(F_\pi^* M_{++}\right)&-&
  \left(1-x+\fr{1}{2} x^2 \right)\, {\rm Re}\left(F_\pi^*
  M_{+-}\right) \Biggr]
  \nonumber
\eea
with $y_2 = W^2/(sx)$.

Analogously, in this limit the muon pair production [see
Eq.~(\ref{38b})] takes the form
\bea
  \varepsilon_+ \varepsilon_-
  \fr{d\sigma_{12}}{ d^3p_+\, d^3p_-}
  &=&  \fr{4\alpha^3}{\pi^3}\, \fr{x}{W^6 d}\,
  \fr{{\bf k}_\perp {\bf \Delta}_\perp}{{\bf k}_\perp^2}\times
  \nonumber \\
  &&\left( 1-y_2 +\fr{1}{ 2}
  y^2_2 \right)\, L_2 \times
  \label{C.5}
  \\
  \left[(1-x) \fr{4\mu^2}{W^2} \right.&-&\left. \left(1-x+\fr{1}{ 2} x^2
  \right) \,\left(2 -
   \fr{{\bf \Delta}^2_\perp}{W^2} \right)
  \right]\,.
  \nonumber
\eea
Both distributions are proportional to the transverse
variable $v={\bf k}_\perp {\bf \Delta}_\perp$
and do not depend on the longitudinal variable
$\xi$. These results coincide with those of
Ref.~\cite{CS}.

\reseteqn
\end{appendix}


\begin{thebibliography}{99}

\bibitem{DAFNE}
  The Second DA$\Phi$NE Physics Handbook, Eds. L.~Maiani,
  G.~Pancheri, N.~Paver (INFN, Frascati, 1995)

\bibitem{Phi}
  Int. Workshop ``$e^+ e^-$ collisions from $\phi$ to $J/\Psi$''
  (Novosibirsk, March 1-5, 1999)

\bibitem{CS}
  V.L.~Chernyak, V.G.~Serbo: Nucl. Phys. {\bf B 67} (1973) 464

\bibitem{Diehl}
  M.~Diehl, T.~Gosset, B.~Pire: hep-ph/0003233

\bibitem{Sav}
  V.~Savinov (CLEO): private communication

\bibitem{Kur}
  M.~Galynsky, E.A.~Kuraev, P.G.~Ratcliffe, B.G.~Shaikhatdenov:
  hep-ph/0003061

\bibitem{BG}
  V.E.~Balakin, V.M.~Budnev, I.F.~Ginzburg:
  Sov. ZhETF Pis'ma {\bf 11} (1970) 559;
  V.M.~Budnev, I.F.~Ginzburg: Phys. Lett. {\bf B 37} (1971) 310

\bibitem{BGMS}
  V.M.~Budnev, I.F.~Ginzburg, G.V.~Meledin, V.G. Serbo:
  Phys. Rep. {\bf 15C} (1975) 575

\bibitem{KS}
  G.L.~Kotkin, V.G.~Serbo:
  Yad. Fiz. {\bf 21} (1975) 785

\bibitem{KLMS}
  E.A.~Kuraev, L.N.~Lipatov, N.P.~Merenkov, M.I.~Strikman:
  Yad. Fiz. {\bf 23} (1976) 163

\bibitem{BFKK}
  V.N.~Baier, V.S.~Fadin, V.S.~Khoze, E.A.~Kuraev:
  Phys. Rep. {\bf 78} (1981) 293

\end{thebibliography}
\end{document}